\documentclass[reprint,amsmath,amsfonts,floatfix]{revtex4-2}

\usepackage{hyperref}  
\hypersetup{%
	pdfstartpage = 1, %
	breaklinks = false, %
	colorlinks = true, %
	linkcolor = black, %
	citecolor = black, %
	filecolor = black, %
	menucolor = black, %
	urlcolor = blue %
}
\usepackage{graphicx}  
\usepackage{multirow}  
\newcommand*{\clozenge}[1]{\includegraphics[scale=0.7]{lozenge#1}}
\newcommand*{\graphicscaling}{0.8}

\usepackage{mathtools}  
\usepackage{mathrsfs}  
\usepackage{amsthm}
\usepackage{thmtools} 
\theoremstyle{theorem}
	\newtheorem{theorem}{Theorem}
	
\theoremstyle{definition}
	\newtheorem{definition}[theorem]{Definition}
	\newcommand*{\defof}[1]{{\textbf{#1}}}
	
\theoremstyle{remark}
	\newtheorem{remark}[theorem]{Remark}
	
\theoremstyle{lemma}
	\newtheorem{lemma}[theorem]{Lemma}
	
\theoremstyle{proposition}
	\newtheorem{proposition}[theorem]{Proposition}

\newcommand*{\upgamma}{\mathrm{\gamma}}
\newcommand*{\upsigma}{\mathrm{\sigma}}
\newcommand*{\bypred}{\;\middle|\;}  
\newcommand*{\e}{\mathrm{e}}  
\newcommand*{\fieldN}[1][]{\mathbb{N}_{#1}}
\newcommand*{\fieldR}[1][]{\mathbb{R}_{#1}}
\DeclareMathOperator{\arsinh}{arsinh}

\DeclareMathOperator{\Ein}{Ein}
\newcommand*{\limp}{\Rightarrow}  
\newcommand*{\liff}{\Leftrightarrow}  
\renewcommand*{\d}[1][]{\mathrm{d}^{#1}\!\!\;{}}  
\newcommand*{\id}[2][]{\;\d[#1]{#2}}  
\newcommand*{\Borelset}{\mathcal{B}}
\DeclareMathOperator{\rank}{rk}
\newcommand*{\argmin}{\mathop{\arg\min}\limits}
\newcommand*{\argmax}{\mathop{\arg\max}\limits}
\newcommand*{\precd}{\mathbin{{\prec}*}}  
\newcommand*{\paths}[1][]{\mathrm{paths}_{#1}}
\DeclareMathOperator{\prm}{prm}
\DeclareMathOperator{\itn}{itn}

\begin{document}

\title{Local structure of sprinkled causal sets}
\author{Christopher J. Fewster}\email{chris.fewster@york.ac.uk}
\author{Eli Hawkins}\email{eli.hawkins@york.ac.uk}
\author{Christoph Minz}\email{cm1757@york.ac.uk}
\author{Kasia Rejzner}\email{kasia.rejzner@york.ac.uk}
\affiliation{Department of Mathematics, Heslington, York YO10 5DD, University of York}
\date{April 26, 2021}

\begin{abstract}
	We describe numerical and analytical investigations of causal sets sprinkled into spacetime manifolds. 
	
	The first part of the paper is a numerical study of finite causal sets sprinkled into Alexandrov subsets of Minkowski spacetime of dimensions $1 + 1$, $1 + 2$ and $1 + 3$.
	In particular we consider the rank 2 past of sprinkled causet events, which is the set of events that are two links to the past. 
	Assigning one of the rank 2 past events as `preferred past' for each event yields a `preferred past structure', which was recently proposed as the basis for a causal set d'Alembertian. 
	We test six criteria for selecting rank 2 past subsets. 
	One criterion performs particularly well at uniquely selecting --- with very high probability --- a preferred past satisfying desirable properties. 

	The second part of the paper concerns (infinite) sprinkled causal sets for general spacetime manifolds. 
	After reviewing the construction of the sprinkling process with the Poisson measure, we consider various specific applications. 
	Among other things, we compute the probability of obtaining a sprinkled causal set of a given isomorphism class by combinatorial means, using a correspondence between causal sets in Alexandrov subsets of $1 + 1$ dimensional Minkowski spacetime and 2D-orders. 
	These methods are also used to compute the expected size of the past infinity as a proportion of the total size of a sprinkled causal set.
\end{abstract}

\maketitle

\clearpage
\tableofcontents

\section{Introduction}
Microscopic phenomena in physics are well described by quantum theory, while the theory of general relativity becomes relevant for the macroscopic regime of gravity. 
The interaction of strong gravitational fields with quantum fields requires a theory of quantum gravity. 
One framework for quantum gravity is causal set theory~\cite{1987BombelliEtAl,2009Henson,2011Sorkin}, which replaces the classical spacetime continuum by the discrete structure of a \emph{causal set} (causet for short) at small length scales. 
One hopes to find the physics of the spacetime continuum at larger length scales emerging from this discrete structure. 

The aim of this paper is twofold. 
On the one hand, it describes a numerical investigation into the local structure of finite causal sets `sprinkled' on Minkowski spacetimes of dimensions $1 + 1$, $1 + 2$, and $1 + 3$. 
On the other hand, it addresses various questions relating to (infinite) causal sets sprinkled on an arbitrary spacetime manifold by analytical means. 

Our investigation is motivated by the problem of describing classical and quantum fields on causets, as a first step towards the larger goal of considering the interaction of the fields with the causets. 
Part of this problem is to find appropriate discrete replacements for the equations of motion for (classical and quantum) fields and associated operators (like the d'Alembertian and its Green's functions)~\cite{2013DowkerGlaser,2014Glaser,2014AslanbeigiSaravaniSorkin}.
A recent approach is based on a new supplementary structure called a \emph{preferred past}~\cite{2020DableheathEtAl}. 
For any causet event, the events that are two links to its past constitute its \emph{rank 2 past}. 
A preferred past structure chooses one of these rank 2 past elements for each causet event, other than those with empty rank 2 past, which are said to belong to the \emph{2-layer past infinity}.

The first main objective of this work is to study selections of the rank 2 past in order to motivate a `good' choice for the preferred past structure by prescribing further conditions on the set of rank 2 past events. 
For this task, we classify the causal intervals $[ x, y ]$ (referred to as \emph{diamonds}) that are spanned by an event $x$ and any of its rank 2 past events $y$. 
We conduct numerical simulations to test 6 criteria that select subsets of the rank 2 past within causets arising from a Poisson process called \emph{sprinkling} that randomly selects a set of events from the given spacetime~\cite{2009Sorkin}, here from Alexandrov subsets of Minkowski spacetime. 
We discuss the statistics of past diamonds that are selected by each criterion. 
Each statistic is based on an ensemble of 10000 sprinkled causets for the flat spacetimes with dimensions $1 + 1$, $1 + 2$ and $1 + 3$. 
Thereby, the dimensional dependence can be visualized. 
As quality indicators for the criteria, we consider the number of rank 2 past events that are selected, the distribution of the selected events projected along the unit hyperboloid (the tendency towards Lorentz invariance), and the proper time separation spanned by the diamonds, with the goal to obtain mostly unique preferred past diamonds with a low cardinality and uniformly distributed along the unit hyperboloid. 
It transpires that one criterion performs particularly well in all the indicators. 

To further study the advantages and disadvantages of the criteria, we present statistics for diamonds that are spanned between next-to-nearest neighbours along all \emph{geodesics} (which are maximal link \emph{paths} for causal sets) connecting the bottom to the top of the sprinkling region (events with the smallest and largest time coordinate). 
Any such diamond along the geodesics only contains $x$, $y$, and events that are linked to $x$ in the past and linked to $y$ in the future; we call it a \emph{pure} diamond. 
Our numerical study shows that the diamonds along the geodesics are very small pure diamonds and their size is almost independent of the dimension of the sprinkled Minkowski spacetime within the range of dimensions we investigated. 

The second main objective concerns (infinite) causal sets sprinkled on a given spacetime manifold. 
We review the rigorous construction of the Poisson probability measure~\cite{1998AlbeverioKondratievRoeckner} and bring it into the context of causal set theory. 
With this, one can compute the probability that a \emph{sprinkle} (a possible outcome of the sprinkling process) belongs to a given causet isomorphism class containing all sprinkles with the same causal relations. 
As an analytically feasible example, we consider an Alexandrov subset in $1 + 1$ dimensional Minkowski spacetime. 
Here, the probability is related to counting all \emph{2D-orders} that correspond to the same causal set. 
The 2D-orders are known to be the product of the total orders of the two null coordinates $( u, v )$ for the sprinkled events~\cite{2008BrightwellHensonSurya}. 
We compare analytically computed probabilities that a uniformly chosen random event of such a sprinkled causet is in the 1-layer or 2-layer past infinities with numerical results. 
Our findings on the 1-layer past infinity confirm previously known results asymptotically for very large sprinkles~\cite{1990Winkler}, while our results on the 2-layer past infinity are new. 
On the one hand, this serves as a consistency check for the numerical techniques, and on the other hand, it demonstrates that the proportion of events without a rank 2 past is negligible for large sprinkles. 

In Sec.~\ref{sec:LocalStructure}, we introduce the notations and terminology for the preferred past structure, so that we can study the diamond to rank 2 past events for finite sprinkles in Sec.~\ref{sec:NumericalResults}. 
The discussion of infinite (sprinkled) causal sets on spacetime manifolds is presented in Sec.~\ref{sec:InfiniteSprinkles}. 
We conclude in Sec.~\ref{sec:Conclusion} and relegate various technical details to the appendices.

\section{(Local) structure of causal sets}
\label{sec:LocalStructure}
In this section, we lay out the necessary notations and definitions to review the \emph{preferred past structure} that was introduced for the discretization of the d'Alembertian in the Klein-Gordon equation on causal sets by~\cite{2020DableheathEtAl}. 
This review leads to a characterization of the \emph{causal intervals} (\emph{diamonds}) that are spanned by events (points of a causal set) and their \emph{rank 2 past}. 

\subsection{Preliminaries}
A causal set is a type of a partially ordered set. 
\begin{definition}
A \defof{partially ordered set} $( \mathscr{C}, \preceq )$ is a set $\mathscr{C}$ equipped with a binary relation $\preceq$ such that the following axioms are fulfilled for all $x, y, z \in \mathscr{C}$ 
\begin{align}
	\label{eq:CausetOrderingReflexivity}
			\text{Reflexivity:} 
	&& 
			x 
	&\preceq x 
	, 
	\\\label{eq:CausetOrderingAntisymmetry}
			\text{Anti-symmetry:} 
	&& 
			( x \preceq y \land y \preceq x ) 
	&\liff x = y 
	, 
	\\\label{eq:CausetOrderingTransitivity}
			\text{Transitivity:} 
	&& 
			( x \preceq y \land y \preceq z ) 
	&\limp x \preceq z 
	. 
\end{align}
If two points $x$ and $y$ are ordered, but not equal $( x \preceq y \land x \neq y )$, we write $x \prec y$. 
\end{definition}
\begin{definition}
Let $( \mathscr{C}, \preceq )$ be a partially ordered set of spacetime events where the partial order $\preceq$ is the causal relation. 
For any pair of events $x, y \in \mathscr{C}$, 
\begin{align}
	\label{eq:CausetCausalIntervalClosed}
			[ x, y ] 
	&:= \left\{ z \in \mathscr{C} \bypred x \preceq z \preceq y \right\} 
\end{align}
defines the \defof{closed causal interval} between $x$ and $y$, while 
\begin{align}
	\label{eq:CausetCausalIntervalOpen}
			( x, y ) 
	&:= \left\{ z \in \mathscr{C} \bypred x \prec z \prec y \right\} 
\end{align}
defines the \defof{open causal interval} between them. Causal intervals are also known as
\defof{Alexandrov sets}. 
A \defof{causal set (causet)} is a partially ordered set $( \mathscr{C}, \preceq )$ that is locally finite, i.e.\ the cardinality of every causal interval (for all $x, y \in \mathscr{C}$) is finite, 
\begin{align}
	\label{eq:CausetLocalFiniteness}
			\text{Local finiteness:} 
	&& 
			\bigl| [ x, y ] \bigr| 
	&< \infty 
	. 
\end{align}
\end{definition}
The axiom of anti-symmetry \eqref{eq:CausetOrderingAntisymmetry} is also referred to as the \emph{axiom of acyclicity} for causal sets, since it ensures that the causet does not have causal loops. 
\begin{figure}
	\centering
	\includegraphics[scale=\graphicscaling]{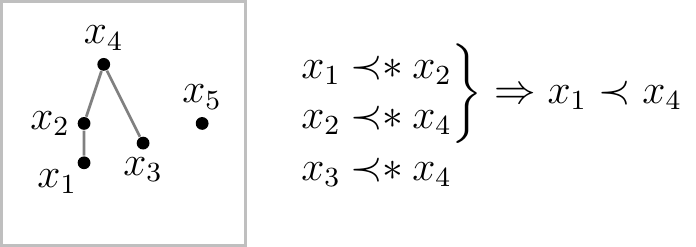}
	\caption{\label{fig:HasseDiagram} Hasse diagram of a causet with 5 events $x_i$ ($i \in \{ 1, 2, 3, 4, 5 \}$) shown as vertices. One can read off the links from the edges of the graph (directed towards the top of the paper). The full causal structure follows by transitivity as we show on the right.}
\end{figure}
\begin{definition}\label{def:Link}
	An event of a causet $( \mathscr{C}, \preceq )$, $x \in \mathscr{C}$ is \defof{linked to} another event $y \in \mathscr{C}$ when $[ x, y ] = \{ x, y \}$ and $x \neq y$. 
	In these circumstances, we write $x \precd y$. 
	A \defof{Hasse diagram} represents a causet as a graph with the events as vertices and the links as edges pointing up the page, see \autoref{fig:HasseDiagram} for an example. 
\end{definition}

Paths in causal sets are concatenations of links.
\begin{definition}
	A \defof{chain} is a totally ordered subset of a causet $\mathscr{C}$. 
	A \defof{path} is a chain such that consecutive events are linked. 
	The set of \defof{paths from $x$ to $y$} is denoted by 
	\begin{align}
		\nonumber
				\paths( x, y ) 
		&:= \bigl\{ 
					\{ x, z_1, z_2, \dotso, z_{n - 2}, y \} \subset [ x, y ] 
				\bigm| 
		\\\label{eq:CausetInvervalPaths}
		&\qquad\quad
					x \precd z_1 \precd \dotsb \precd z_{n - 2} \precd y  
				\bigr\}
		. 
	\end{align}
	We call a path from $x$ to $y$ \defof{minimal} (resp., \defof{maximal}) if it has minimal (resp., maximal) cardinality among the elements of $\paths( x, y )$. 
\end{definition}
In particular, a maximal path is a causet analogue of a timelike geodesic in the continuum (see also Sec.~\ref{sec:GeodesicPath}). 

\subsection{Past and future of causal set events}
The past and future of events and subsets of events in a causal set are defined by analogy with the continuum. The following conventions agree with those given in~\cite{2020DableheathEtAl}.
\begin{definition}
	The \defof{past} ($-$) and \defof{future} ($+$) of an event $x \in \mathscr{C}$ or subset $\mathscr{D} \subset \mathscr{C}$ in the causet $\mathscr{C}$ are given by
	\begin{subequations}
	\begin{align}
		\label{eq:CausetPast}
				J^{-}( x ) 
		&:= \left\{ 
					y \in \mathscr{C} 
				\bypred 
					y \preceq x 
				\right\} 
		, 
		\\\label{eq:CausetFuture}
				J^{+}( x ) 
		&:= \left\{ 
					y \in \mathscr{C} 
				\bypred 
					x \preceq y 
				\right\} 
		, 
		\\\label{eq:CausetPastFutureSets}
				J^{\mp}( \mathscr{D} ) 
		&:= \bigcup_{x \in \mathscr{D}} J^{\mp}( x ) 
		. 
	\end{align}
	\end{subequations}
\end{definition}
The past and future of a point can be partitioned into \emph{layers} and \emph{ranks}.
\begin{definition}
	The \defof{layer $k$ past and future} of a point $x \in \mathscr{C}$ in the causet $\mathscr{C}$ are the sets 
	\begin{subequations}
	\begin{align}
		\label{eq:CausetPastLayers}
				L^{-}_{k}( x ) 
		&:= \left\{ 
					y \in J^{-}( x ) 
				\bypred 
					\bigl| [ y, x ] \bigr| - 1 = k 
				\right\} 
		, 
		\\\label{eq:CausetFutureLayers}
				L^{+}_{k}( x ) 
		&:= \left\{ 
					y \in J^{+}( x ) 
				\bypred 
					\bigl| [ x, y ] \bigr| - 1 = k 
				\right\} 
		, 
	\end{align}
	\end{subequations}
	respectively, where $k \in \fieldN[0]$~\cite{2009Sorkin}. 
\end{definition}
\begin{definition}
	The \defof{$k$-layer past (or future) infinity} of a causet $\mathscr{C}$ is the set 
	\begin{align}
		\label{eq:CausetPastFutureInfinityLayers}
				C^{\mp}_{k} 
		&:= \left\{ 
					x \in \mathscr{C} 
				\bypred 
					\forall j \geq k: 
					L^{\mp}_{j}( x ) = \emptyset 
				\right\} 
		. 
	\end{align}
\end{definition}
\begin{definition}
	Given a causet $\mathscr{C}$, the \defof{rank} of an event $y \in \mathscr{C}$ relative to another event $x \in \mathscr{C}$ is 
	\begin{align}
		\label{eq:CausetRank}
				\rank( y, x ) 
		&:= \begin{dcases}
					\min_{\mathscr{P} \in \paths( x, y )} | \mathscr{P} | - 1 
				& x \preceq y 
				, \\
					\infty 
				& \text{otherwise} 
				. 
				\end{dcases}
	\end{align}
	
\end{definition}
Notice that for every element $x$ of a causet $\mathscr{C}$: $\paths( x, x ) = \{ \{ x\} \}$, so that it is in the zeroth rank to itself, $\rank( x, x ) = 0$. 
The relative rank of two spacelike separated events is infinite. The future and past of an event may be partitioned by rank.
\begin{definition}
	The \defof{rank $k$ past and future} of an event $x \in \mathscr{C}$ in the causet $\mathscr{C}$ are the sets 
	\begin{subequations}
	\begin{align}
		\label{eq:CausetPastDiamonds}
				R^{-}_{k}( x ) 
		&:= \left\{
					y \in J^{-}( x ) 
				\bypred
					\rank( x, y ) = k 
				\right\}
		, 
		\\\label{eq:CausetFutureDiamonds}
				R^{+}_{k}( x ) 
		&:= \left\{
					y \in J^{+}( x ) 
				\bypred
					\rank( y, x ) = k 
				\right\}
		, 
	\end{align}
	\end{subequations}
	respectively, where $k \in \fieldN[0]$. 
\end{definition}

The classification of points by layer or rank plays an important role in the definition of discretized wave operators on casual sets. 
These discretizations typically involve a weighted sum taken over field values with weights determined by the layer or rank relative to the point where the operator is notionally evaluated.
For example, the discretizations studied in~\cite{2009Sorkin,2013DowkerGlaser,2014Glaser,2014AslanbeigiSaravaniSorkin} take a different number of layers into account depending on the spacetime dimension that is described by the causal set. 
The spacetime dimension is not a pre-defined property of a causet, but has to be estimated by the Myrheim-Meyer estimator~\cite{1978Myrheim,1988Meyer} or other approximations~\cite{2003Reid,2013RoySinhaSurya}. 
A more recent alternative approach~\cite{2020DableheathEtAl} proposes a discretization scheme for the wave operators that, while taking its inspiration from a discrete lattice causet in $1+1$ dimensions, has the aim of being dimension-independent. 
Although this approach does not need the approximated spacetime dimension as an input, it does require the specification of an additional \emph{preferred past structure}. 
One of the primary goals of this paper is to investigate ways in which a preferred past may be associated intrinsically to a causal set and to evaluate their performance on sprinkled causets in Minkowski spacetime.

\subsection{Preferred past structure and diamonds}
\label{subsec:Diamonds}
\begin{figure*}
	\centering
	\includegraphics[scale=\graphicscaling]{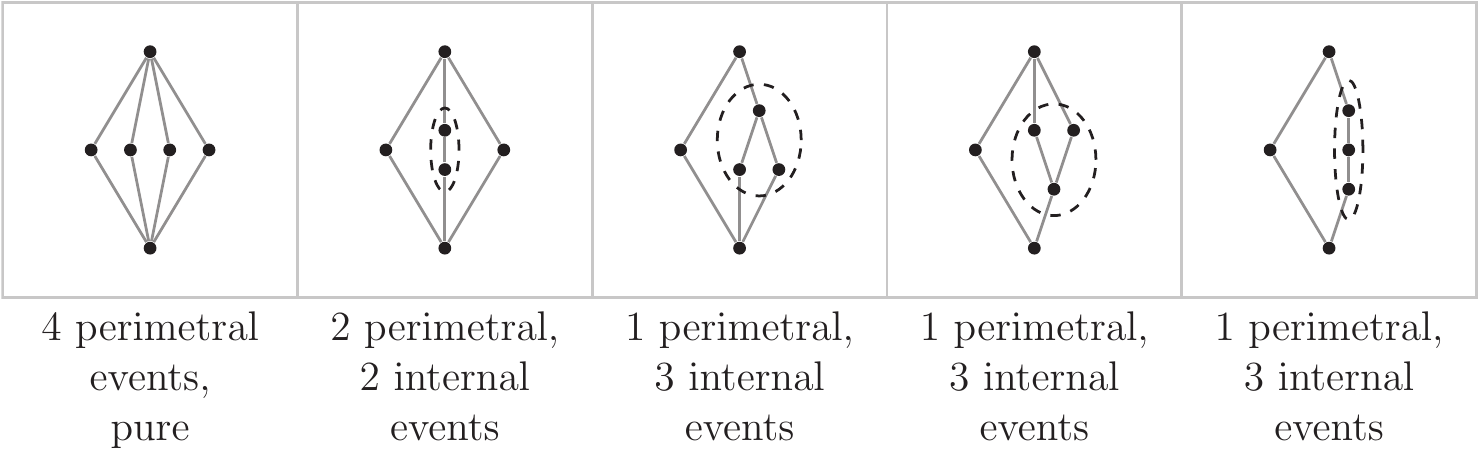}
	\caption{\label{fig:4DiamondClassification} Types of 4-diamonds $[ x, y ]$ spanned from events $x$ to $y \in R^{+}_{2}( x )$. From left to right, the pure 4-diamond, followed by the 4-diamond with 2 internal events (dashed ellipse), and three 4-diamonds with 3 internal events.}
\end{figure*}
The rank 2 past (and future) is the basis for the \emph{preferred past structure}. 
\begin{definition}\label{def:PreferredPastStructure}
	Given a causet $\mathscr{C}$, a \defof{preferred past (or future) structure}~\cite{2020DableheathEtAl} is a map 
	\begin{align}
		\label{eq:CausetPreferredPastFutureStructureMap}
				\varLambda^{\mp} : \mathscr{C} \setminus C^{\mp}_{2} 
		&\to \mathscr{C} 
		, 
	\end{align}
	such that 
	\begin{align}
		\label{eq:CausetPreferredPastFutureStructure}
				\varLambda^{\mp}( x ) 
		& \in R^{\mp}_{2}( x ) 
	\end{align}
	holds for all events $x$ that are not in the 2-layer past (or future) infinity, i.e.\ $x \in \mathscr{C} \setminus C^{\mp}_{2}$. 
\end{definition}

The discretized Klein--Gordon operator $\Box$ proposed in~\cite{2020DableheathEtAl} is defined as follows. 
Suppose a preferred past $\varLambda^{-}$ is specified on a causet $\mathscr{C}$ and let $\phi : \mathscr{C} \to \fieldR$ be a scalar field. Then $\Box \phi : \mathscr{C} \setminus C^{-}_{2} \to \fieldR$ is defined at $x \in \mathscr{C} \setminus C^{-}_{2}$ as a weighted sum over the values of $\phi$ on $[ \varLambda^{-}( x ), x ]$, 
\begin{align}
	\label{eq:CausetDAlembertian}
			( \Box \phi )( x ) 
	&= \phi( \varLambda^{-}( x ) ) 
			- \frac{2}{| I_x |} 
				\sum_{z \in I_x} \phi( z ) 
			+ \phi( x ) 
	,
\end{align}
where $I_x = ( \varLambda^{-}( x ), x )$ is the open causal interval from the preferred past of $x$. 
For a full discussion of this discretization method, see~\cite{2020DableheathEtAl}. 
In this paper, we will focus our attention on the preferred past structure itself. 
For an arbitrary causet event $x \in \mathscr{C}$, the rank 2 past $R^{-}_{2}( x )$ generally contains more than one event. 
Below, we analyse 6 methods for selecting subsets of $R^{-}_{2}( x )$ with the aim to find one method that (among other things) selects singleton sets with high probability. 
To this end, we introduce further properties of the open causal intervals spanned between an event and any of the events in its rank 2 past, and also of the events within such intervals. 
\begin{definition}
	For a causet $\mathscr{C}$ and a pair of events $x, y \in \mathscr{C}$ such that $\rank( y, x ) = 2$, 
	we call the Alexandrov set $[ x, y ]$ a \defof{diamond} with \defof{diamond size} given by the cardinality of its open interval,  
	\begin{align}
		\label{eq:CausetDiamondSize}
				k
		&= \bigl| ( x, y ) \bigr| 
		. 
	\end{align}
	It is a \defof{past $k$-diamond} of the event $y$ and a \defof{future $k$-diamond} of the event $x$. 
	In particular, we call it the \defof{preferred past (future) diamond} if the event $x$ (or $y$) is the preferred past (future) of $y$ (of $x$) with respect to some preferred past (future) structure. 
\end{definition}
Events in the open interval $( x, y )$ can either be only linked to $x$ and $y$, or they are related to other events in this set, which leads to the following diamond properties. 
\begin{definition}\label{def:DiamondProperties}
	Let $[ x, y ]$ be a $k$-diamond in the causet $\mathscr{C}$. 
	We call an event $z \in ( x, y )$ \defof{perimetral} if $x \precd z \precd y$, so that the number of perimetral events is 
	\begin{align}
		\label{eq:CausetDiamondPerimetral}
				\prm( y, x ) 
		&:= \bigl| \{ z \in ( x, y ) \;|\; x \precd z \precd y \} \bigr| 
		. 
	\end{align}
	We call an event $z \in ( x, y )$ \defof{internal} if it is not perimetral. 
	There are 
	\begin{align}
		\nonumber
				\itn( y, x ) 
		&:= \bigl| ( x, y ) \bigr| - \prm( y, x ) 
		\\\label{eq:CausetDiamondInternal}
		&= k - \prm( y, x ) 
	\end{align}
	internal events in $( x, y )$. 
	We call the diamond \defof{pure} if $\mathrm{itn}( y, x ) = 0$. 
\end{definition}
Notice that the number of perimetral events of a diamond \eqref{eq:CausetDiamondPerimetral} is the same as the number of minimal paths. 
As an example, consider two events $x, y \in \mathscr{C}$ such that their interval $[ x, y ]$ is a 4-diamond. 
For it to be a 4-diamond, there has to be at least one event $z$ in a pure relation $x \precd z \precd y$, but the remaining three events can have an arbitrary causal arrangement, so there are the 5 distinct 4-diamonds drawn in \autoref{fig:4DiamondClassification}. 

\section{Numerical results for sprinklings on flat spacetime}
\label{sec:NumericalResults}
As we discussed in the previous section, diamonds are spanned between events of a causet and their rank 2 past events. 
The main aim of the simulations is to analyse how to reduce the choices for the preferred past structure of a causet by choosing subsets of the rank 2 past for the causet events that are singletons, at least with high probability. 
We carry out the investigation for three flat spacetimes with dimensions from $1 + 1$ to $1 + 3$ so that the dimensional dependence can be studied. 

\subsection{Outline of the simulations}
We conducted the simulations with MATLAB R2018a code and utilized the Viking high performance computing cluster of the University of York. 

For each dimension $d = 1 + 1$, $1 + 2$ and $1 + 3$ of a Minkowski spacetime $\mathbb{M}^{d}$, we consider a non-empty Alexandrov subset $U = J^{+}( p ) \cap J^{-}( q )$ for fixed $p, q$ in $\mathbb{M}^{d}$. 
On the subset $U$, we repeat a sprinkling process 10000 times with a fixed sprinkling density parameter such that the sprinkles have an expected cardinality of 6000 events. 
This corresponds to a grand-canonical ensemble of sprinkles in the given Alexandrov set. 
For each event $x$ in each sprinkled causet, we consider every event $y \in R^{-}_{2}( x )$ in the rank 2 past of $x$ and count the number of perimetral and internal events in the diamonds spanned by $x$ and $y$. 
The counts are accumulated over all the 10000 sprinkles so that we obtain results averaging over tens of millions of rank 2 past events. 
Details on the implementation of the sprinkling process are given in Appendix~\ref{sec:AppendixSprinklingImplementation}. 

\begin{figure}
	\centering
	\includegraphics[scale=0.65]{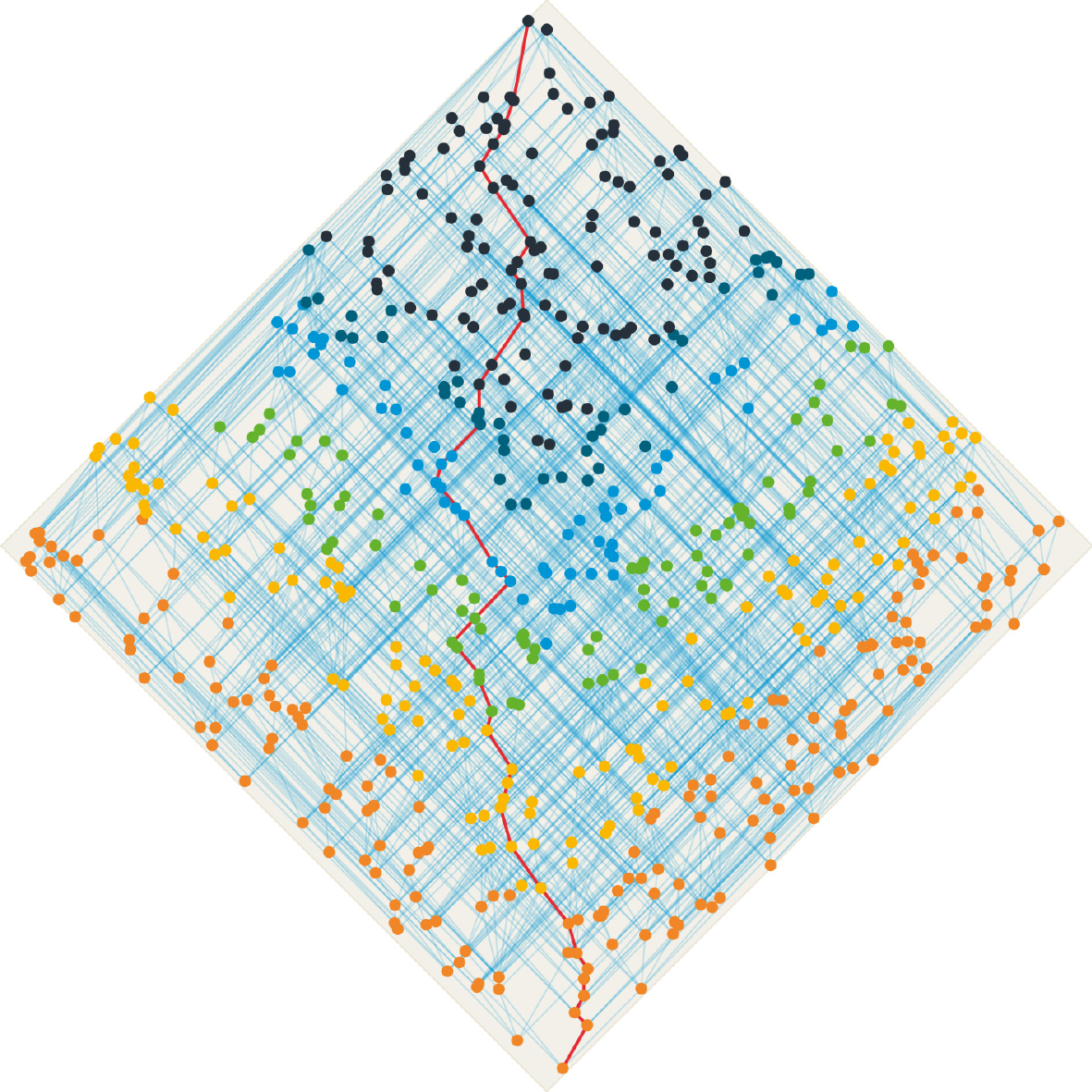}
	\caption{\label{fig:VolReductionGeodesicPath} Example sprinkle of 600 events into an Alexandrov subset of $1 + 1$ dimensional flat spacetime. To analyse the effects of the past infinity, we observe the events that are sprinkled in a reduced region $U_{i}$, from the entire region $i = 0$ (no reduction) up to $i = 5$ (smallest observation region, black shade). One possible maximal path (\emph{timelike geodesic}, see Sec.~\ref{sec:GeodesicPath}) is shown by a thick (red) line that connects the past-most to the future-most event.}
\end{figure}
There will be effects from the past boundary of the sprinkling region $U$. 
To mitigate these effects, we set up various \emph{observation regions} as subsets of $U$. 
For $i \in \{ 0, 1, \ldots, 5 \}$, fix points $p_i \in \mathbb{M}^{d}$ along the straight line from $p$ to $q$ in $\mathbb{M}^{d}$ such that the observation regions $U_{i} = J^{+}( p_i ) \cap J^{-}( q )$ have a volume 
\begin{align}
	\label{eq:VolReductionFactors}
			V_{i} 
	&= 2^{-d i / 4} V_{0} 
	. 
\end{align}
So we obtain 6 regions per Alexandrov subset of Minkowski space to compare. 
We consider all rank 2 past diamonds in $U$ whose future tip is contained in $U_{i} \subset U$. 
In \autoref{fig:VolReductionGeodesicPath}, for example, the observation region with volume $V_{1}$ excludes the lowest events (orange), $V_{2}$ further excludes the next set of events (yellow), then $V_{3}$ also excludes the darker shaded events (green), and so on. 

In Sec.~\ref{sec:Rank2PastCriteria} we set out $6$ methods for selecting a subset $S( x ) \subset R^{-}_{2}( x )$ for each event $x$ outside the rank $2$ past infinity. 
The first criterion was proposed in~\cite{2020DableheathEtAl}, while the others are newly introduced here. 
We compare the subsets $S(x)$ selected by each method so that we can identify the one that performs best in relation to three qualitative measures:
\begin{itemize}
	\item the selected sets $S(x)$ should be singletons with high probability, across all points $x$ in each sprinkle in the ensemble studied
	\item the distribution of proper time separations between $x$ and the event(s) in $S(x)$ should have low variance and small expectation value, across the ensemble as before
	\item the distribution of the unit-normalized separation vectors between $x$ and the event(s) in $S(x)$, should be approximately uniformly distributed on the unit hyperboloid, across the ensemble.
\end{itemize}
The third of these is intended to ensure Lorentz invariance of the preferred past structure, in a statistical sense, in the limit of large sprinkles.

Furthermore, we study the diamond size and its expected proper time separation in more generality. 
Consider the events with the minimal and maximal time coordinates in a given sprinkle. 
If they are causally related, as occurs with high probability, there are maximal paths between them; 
an example is illustrated as the line connecting the events with the smallest and largest time coordinate in \autoref{fig:VolReductionGeodesicPath}. 
Such paths are analogous to timelike geodesics and may be regarded as potential observer trajectories. We compute the expected diamond size and proper time separation between next-to-nearest neighbours along such paths.
It transpires that an observer travelling along such a path can hardly determine the dimension of the underlying flat spacetime by measuring the diamond size or the expected proper time separation (ticking rate of a `diamond clock') of the diamonds spanned between next-to-nearest neighbours along the geodesic, see Sec.~\ref{sec:GeodesicPath}.

\subsection{Criteria for selecting rank 2 past subsets}
\label{sec:Rank2PastCriteria}
As described in \autoref{def:PreferredPastStructure}, a preferred past structure maps each causet event outside the 2-layer past infinity to one of their rank 2 past events. 
In general, for any event $x \in \mathscr{C} \setminus C^{-}_{2}$, the rank 2 past $R^{-}_{2}( x )$ contains multiple events $y$ (see also Appendix~\ref{sec:AppendixRank2Past}). 
The diamonds $[ y, x ]$ can be grouped by their number of minimal paths $\prm( x, y )$ and their number of internal events $\itn( x, y )$ as given in \autoref{def:DiamondProperties}. 
We introduce 6 criteria that select events in the rank 2 past whose corresponding diamonds have a specific size (and a specific number of internal events). 
We evaluate these selection criteria against the desirable features described above. 

To begin, we define some notation. 
For any causet event $x \in \mathscr{C} \setminus C^{-}_{2}$, let 
\begin{align}
	\label{eq:DiamondsMaxPerimetral}
			D^{-}_{\max\prm}( x ) 
	&:= \argmax_{y \in R^{-}_{2}( x )} \prm( x, y ) 
\end{align}
denote the set of events $y$ in the past of $x$ that span diamonds with maximal number of perimetral events. 
Here, $\argmax$ (and similarly $\argmin$) of a function yields the set of points of the function domain, where the function becomes maximal (or minimal, respectively). 
Furthermore, let 
\begin{align}
	\label{eq:DiamondsPure}
			D^{-}_{\mathrm{pure}}( x ) 
	&:= \left\{ 
			y \in R^{-}_{2}( x ) 
		\bypred 
			\itn( x, y ) = 0 
		\right\}
\end{align}
be the set of events spanning pure diamonds only. 
We now set out the six criteria that are compared in our simulations. 
For $i \in \{ 1, 2, \ldots, 6 \}$, rule $i$ selects a subset $D^{-}_{\mathrm{crit\,}i}( x ) \subset R^{-}_{2}( x )$ of the rank 2 past of each event $x \in \mathscr{C} \setminus C^{-}_{2}$. 

\begin{figure}
	\centering
	\includegraphics[scale=\graphicscaling]{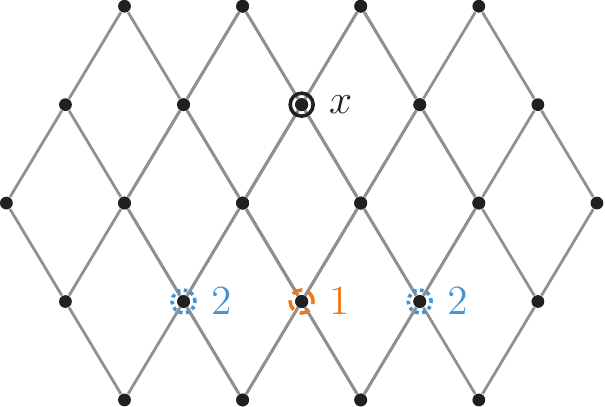}
	\caption{\label{fig:LatticeCausetD2} Subset of a regular 2-dimensional lattice with an element $x$ and its 3 rank 2 past events. The event labelled by 1 is the preferred past corresponding to the largest past diamond of $x$, while the 2 events labelled by 2 correspond to smallest past diamonds of $x$.}
\end{figure}
\begin{figure}
	\centering
	\includegraphics[scale=\graphicscaling]{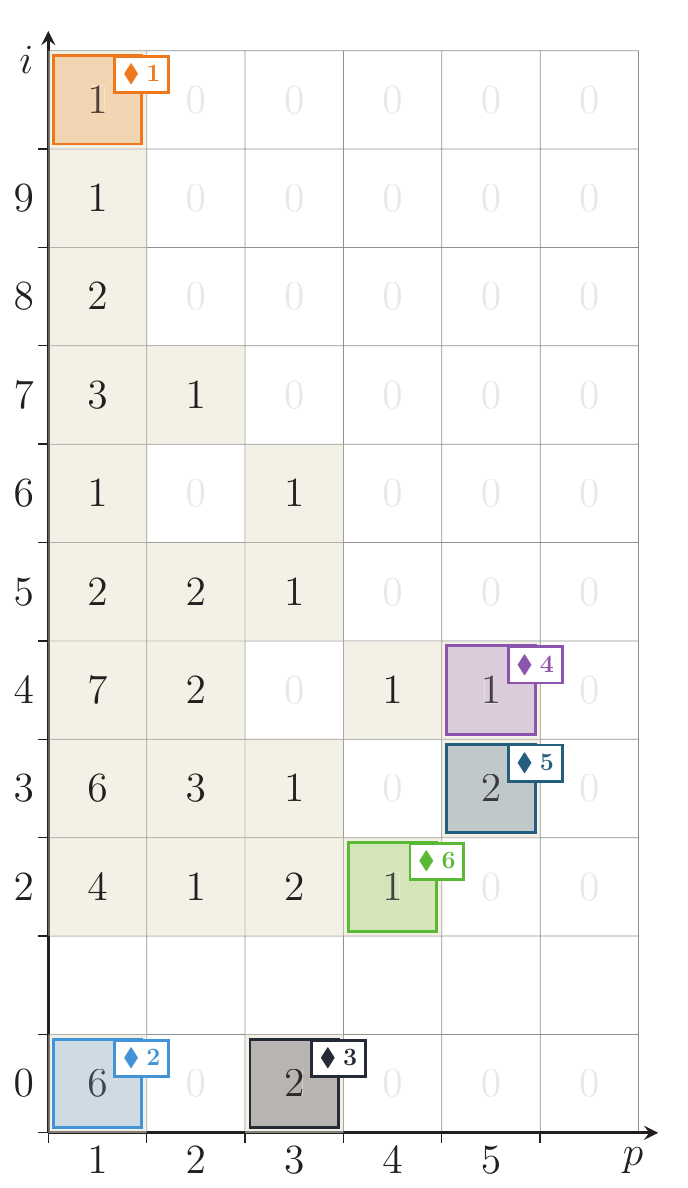}
	\caption{\label{fig:DiamondCountMatrix} Example matrix of the number of past diamonds for an event in a sprinkled causet to demonstrate the 6 criteria. The $p$-axis labels the number of perimetral points and the $i$-axis represents the number of internal events (note that diamonds with 1 internal event do not exist). Thus, in this example there are 7 diamonds with 1 perimetral and 4 internal events. The selection for each of the 6 criteria is labeled. Criterion 6, in particular, picks one of the singletons (which are the ones in this matrix).}
\end{figure}
To motivate our first criterion, consider a regular lattice as depicted in \autoref{fig:LatticeCausetD2}, which has an obvious choice of a preferred past for every element $x$ characterized as the largest past diamond corresponding to the event labelled by 1. 
Using this preferred past structure with \eqref{eq:CausetDAlembertian} yields a good approximation to the d'Alembertian in the continuum limit \cite{2020DableheathEtAl}. 
So the first criterion comprising those $y \in R^{-}_{2}( x )$ such that the diamond $[ y, x ]$ is one of the 
\begin{description}
	\item[\clozenge{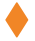} 1] 
		largest diamonds (``maximal layer rule'' proposed in~\cite{2020DableheathEtAl}), 
		\begin{align}
			\label{eq:DiamondsCrit1}
					D^{-}_{\mathrm{crit\,1}}( x ) 
			&:= \argmax_{y \in R^{-}_{2}( x )} \bigl| [ y, x ] \bigr| 
			. 
		\end{align}
\end{description}
For the sprinkled causets, it will turn out that choosing the largest diamond is not the best criterion, since it tends to yield a very large proper time separation between $x$ and $y$ that is only limited by the finite past cardinality of $x$ in our simulation. 
In order to get the smallest proper time, we consider the 
\begin{description}
	\item[\clozenge{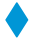} 2] 
		smallest diamonds, 
		\begin{align}
			\label{eq:DiamondsCrit2}
					D^{-}_{\mathrm{crit\,2}}( x ) 
			&:= \argmin_{y \in R^{-}_{2}( x )} \bigl| [ y, x ] \bigr| 
			. 
		\end{align}
\end{description}
We will see that these diamonds correspond to the smallest proper time separation, but they are not unique for the regular lattice nor for typical events in a sprinkled causet. 
In further criteria, we consider maximizing and minimizing the diamond properties of the number of internal and perimetral events. 
Physically, perimetral events of a diamond $[ y, x ]$ in the sprinkle are points that fall very close to the boundary of the Alexandrov subset from $y$ to $x$ within $\mathbb{M}^{d}$, while internal events form time-like paths between these two events. 
As the d'Alembertian describes the propagation of light, we want to maximize the number of perimetral events, so we compare the 
\begin{description}
	\item[\clozenge{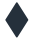} 3] 
		largest (or maximal perimetral) pure diamonds, 
		\begin{align}
			\nonumber
					D^{-}_{\mathrm{crit\,3}}( x ) 
			&:= \argmax_{y \in D^{-}_{\mathrm{pure}}( x )} \bigl| [ y, x ] \bigr| 
			\\\label{eq:DiamondsCrit3}
			&= \argmax_{y \in D^{-}_{\mathrm{pure}}( x )} \prm( x, y ) 
			;
		\end{align}
	\item[\clozenge{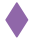} 4] 
		diamonds with the most internal events among the diamonds with the most perimetral events,
		\begin{align}
			\label{eq:DiamondsCrit4}
					D^{-}_{\mathrm{crit\,4}}( x ) 
			&:= \argmax_{y \in D^{-}_{\max\prm}( x )} \itn( x, y ) 
			;
		\end{align}
	\item[\clozenge{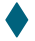} 5] 
		diamonds with the least internal events among the diamonds with the most perimetral events, 
		\begin{align}
			\label{eq:DiamondsCrit5}
					D^{-}_{\mathrm{crit\,5}}( x ) 
			&:= \argmin_{y \in D^{-}_{\max\prm}( x )}{\itn( x, y )} 
			. 
		\end{align}
\end{description}
It might be expected that criterion 4 does not perform the best as it yields diamonds that may also contain a larger number of internal events. 
This presumption will be supported by the comparison of the results for criteria 3 to 5. 
Criteria 3 and 5 can still be refined and we suggest one possible improvement, which will give even better results. 
The 6th criterion is designed to combine the best features of criteria 3 and 5. 
Our results will show that criterion 5 selects a single rank 2 past event with high probability, but its proper time distribution has a large variance. 
On the other hand, criterion 3 yields a prominent peak for the proper time separation, but with a lower probability of selecting a singleton. 
This suggests the following rule:  
\begin{description}
	\item[\clozenge{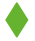} 6] 
		Select the same subset as criterion 5 when there are no singletons among the sets of rank 2 past events $y$ so that $[ y, x ]$ contains $i \in \fieldN[0]$ internal events and $p \in \fieldN$ perimetral events, 
		\begin{align}
			\nonumber
					D^{-}_{i, p}( x ) 
			:= \bigl\{ 
					y \in R^{-}_{2}( x ) 
				\bigm| 
					\itn( x, y ) &= i, 
			\\\label{eq:DiamondsSetMatrix}
					\prm( x, y ) &= p 
				\bigr\} 
			. 
		\end{align}
		If there is at least one singleton among \eqref{eq:DiamondsSetMatrix}, then choose the singleton with the indices 
		\begin{align}
			\label{eq:DiamondsCrit6Condition1}
					j( x ) 
			&:= \min\Bigl\{ i \in \fieldN[0] 
					\Bigm| 
						\bigl| D^{-}_{i, p}( x ) \bigr| = 1 
					\Bigr\} 
			, \\\label{eq:DiamondsCrit6Condition2}
					q( x ) 
			&:= \max\Bigl\{ p \in \fieldN 
					\Bigm| 
						\bigl| D^{-}_{j( x ), p}( x ) \bigr| = 1 
					\Bigr\} 
		\end{align}
		to minimize the number of internal events first and then maximize the number of perimetral events. 
		So 
		\begin{align}
			\label{eq:DiamondsCrit6}
					D^{-}_{\mathrm{crit\,6}}( x ) 
			&:= \begin{cases}
						D^{-}_{\mathrm{crit\,5}}( x ), 
					& \text{if no singletons}, \\ 
						D^{-}_{j( x ), q( x )}( x ), 
					& \text{if $j(x) < \infty$}
					\end{cases}
		\end{align}
\end{description}
The 6th criterion yields events that correspond to diamonds with a size between the size of the diamonds selected by criteria 3 and 5. 
If criterion 3 selects a singleton, criterion 6 selects the same singleton. 
The subset selected by criterion 6 is only non-singleton if there is no singleton among all the subsets \eqref{eq:DiamondsSetMatrix}, so that it selects the same subset as criterion 5. 
Note that this list of criteria is not exhaustive and one might consider further criteria determined by other diamond properties.

All criteria yield non-empty subsets of the rank 2 past for an event outside the 2-layer past infinity, see \autoref{fig:DiamondCountMatrix} for an example and Appendix~\ref{sec:AppendixRank2PastCriteria} for the proofs. 
Notice that similar criteria could be considered for subsets of the rank 2 future $R^{+}_{2}( x )$ for any causet event $x \in \mathscr{C} \setminus C^{+}_{2}$. 
The statistics for rank 2 future subsets are equivalent to the statistics for rank 2 past subsets, because of the time symmetry for the Alexandrov subsets of Minkowski spacetimes. 

\subsection{Cardinality of the rank 2 past subsets}
\label{sec:Rank2PastCriteriaUniqueness}
\begin{figure}
	\centering
	\includegraphics[scale=\graphicscaling]{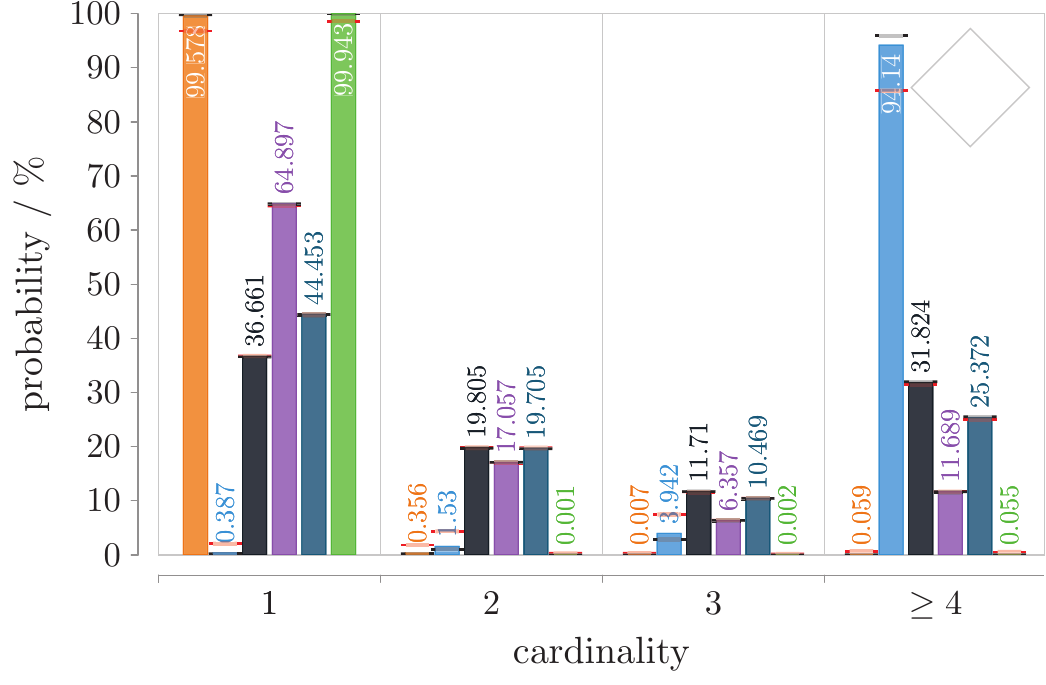}
	\includegraphics[scale=\graphicscaling]{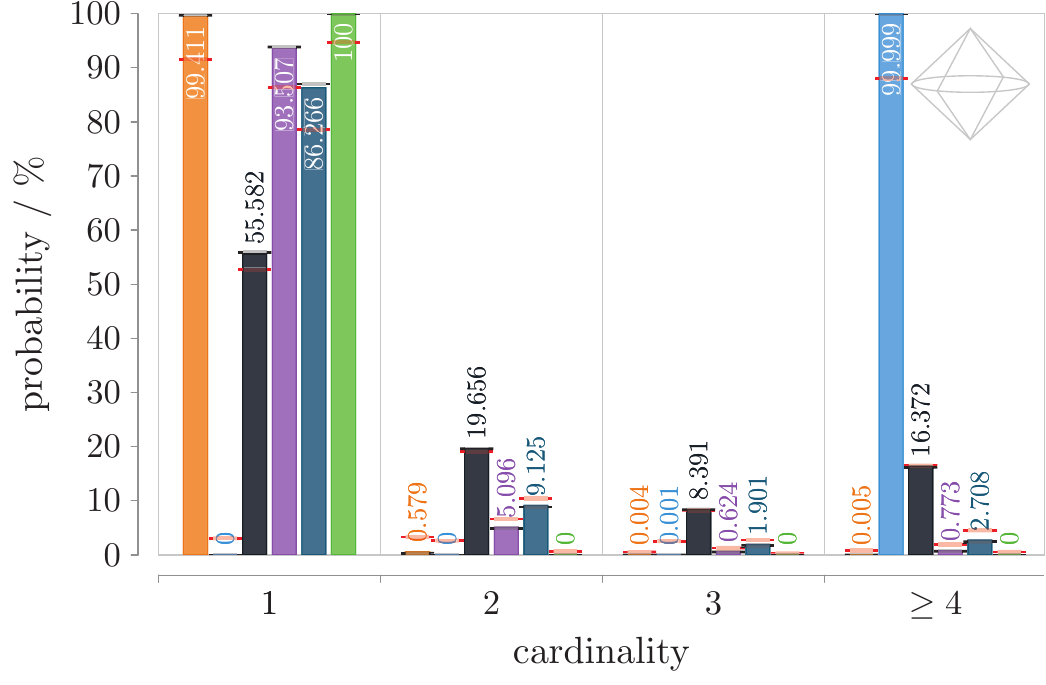}
	\includegraphics[scale=\graphicscaling]{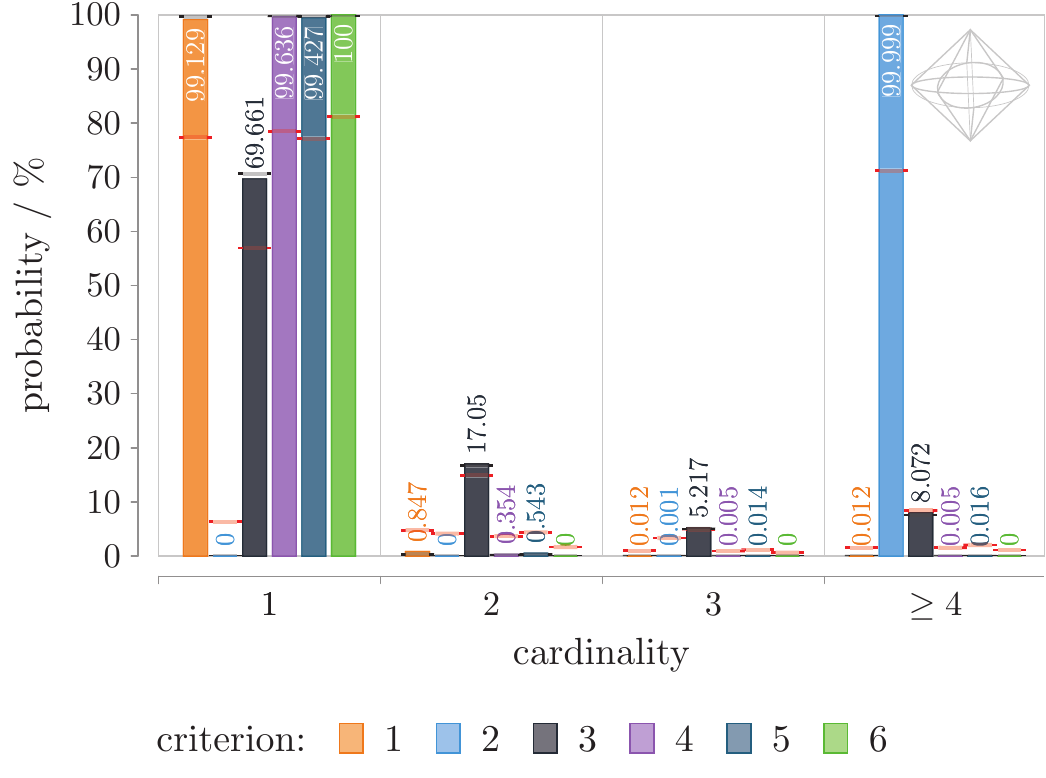}
	\caption{\label{fig:PrefPasts} Discrete probability distributions for the cardinality of the subset of rank 2 past events that are selected by the 6 criteria, observed for the region $U_{2}$ in sprinkles with an expected cardinality of 6000. The horizontal lines across each column indicate the value obtained using the entire sprinkling region $U_{0}$ in red and the smallest observation region $U_{5}$ in black.}
\end{figure}
\autoref{fig:PrefPasts} displays the size distribution of the sets selected by each criterion, in dimensions $1 + 1$ (top), $1 + 2$ (middle) and $1 + 3$ (bottom) as indicated by outlines in the diagrams' top right corners, using observation region $U_{2}$ to mitigate edge effects (see \eqref{eq:VolReductionFactors}). 
To indicate how the results depend on the observation region, each bar is accompanied by horizontal red and black lines corresponding to the values that would be obtained if observing the entire sprinkling region $U_{0}$, or the smallest region $U_{5}$, respectively. 
Note that the latter deviates less from the bar than the former, indicating that edge effects are substantially ameliorated when using $U_{2}$, even though the influence of the past infinity increases with dimension. 

The probability of selecting a singleton (unique rank 2 past event) increases with the spacetime dimension to almost certainty at dimension $1 + 3$ for all criteria but the 2nd and 3rd criterion. 
The 3rd criterion, selecting the rank 2 past events associated to the largest pure diamonds, also shows an increase in the probability for a unique preferred past with increasing spacetime dimension, but for about 30\% of the events there is still more than one rank 2 past event selected at dimension $1 + 3$. 
The 2nd criterion selects mostly the 1-diamonds that are formed by a single 3-path (smallest possible diamond), so that the number of rank 2 past events is very large and, furthermore, increases with the spacetime dimension. 
The 1st criterion performs very well across our results for all 3 dimensions. 
Criterion 6 selects a singleton if and only if there is at least one singleton among all the subsets in matrix \eqref{eq:DiamondsSetMatrix}. 
The chance to find an event in region $V_{2}$ for which criterion 6 selects more than one rank 2 past event is almost as low as 1 in 925000 for dimension $1 + 2$. 
For the observation region $U_{2}$ in dimension $1 + 3$, criterion 6 selects a singleton with certainty within our numerical accuracy, so that the probability to find a non-singleton is less than $10^{-7}$. 

Criterion 6 has the highest probability to select a unique rank 2 past event, followed in order by criteria 1, 4, 5, 3, 2, where criteria 1, 4, and 5 are equally good at dimension $1 + 3$. 

\subsection{Proper time separation for the rank 2 past subsets}
\label{sec:Rank2PastCriteriaProperTimes}
\begin{figure}
	\centering
	\includegraphics[scale=\graphicscaling]{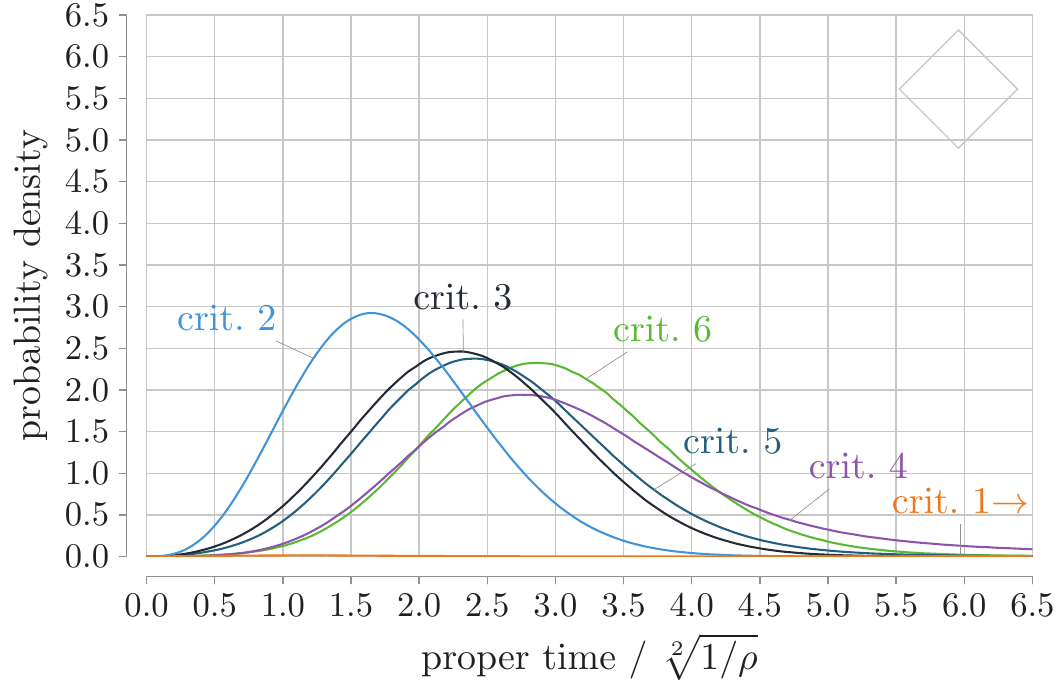}
	\includegraphics[scale=\graphicscaling]{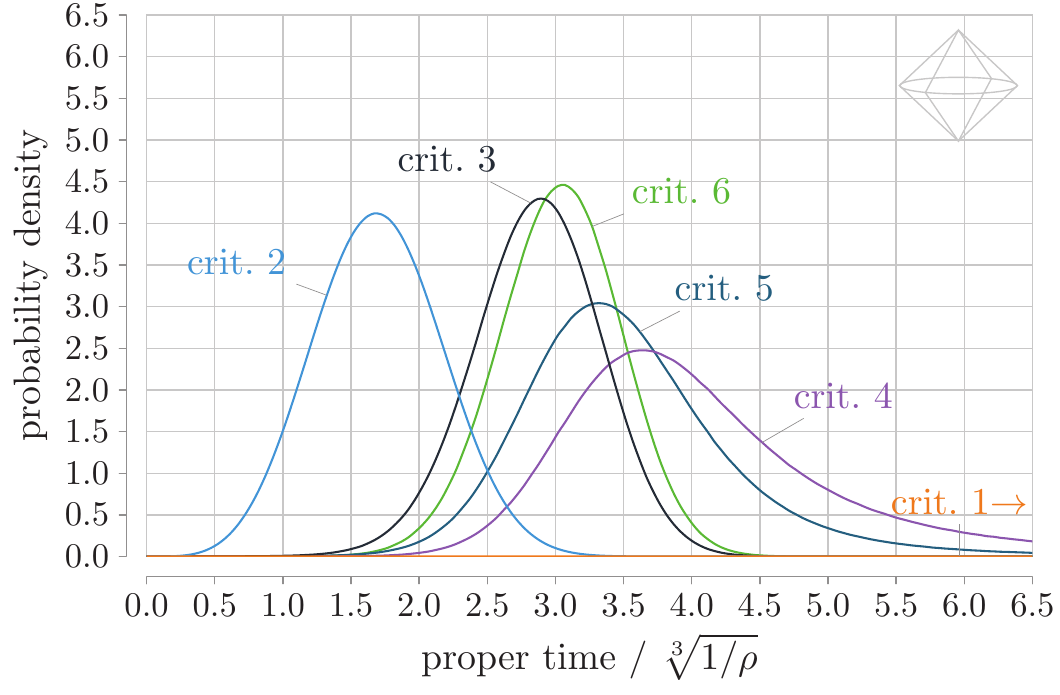}
	\includegraphics[scale=\graphicscaling]{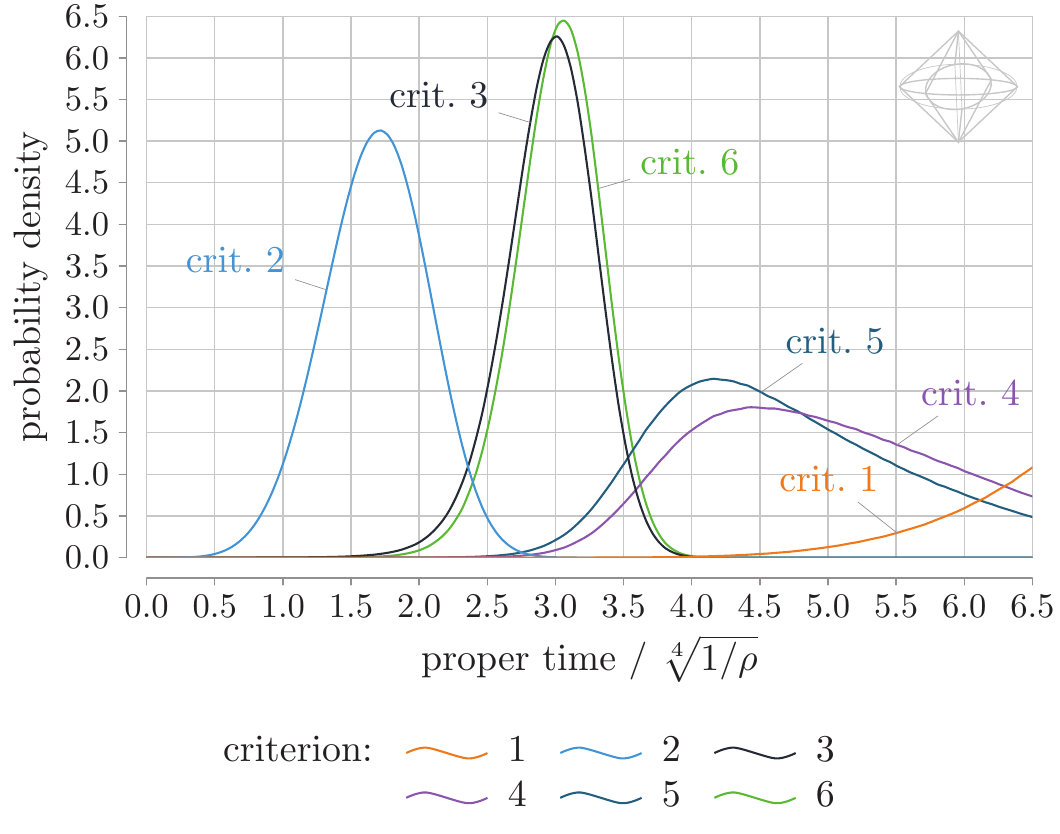}
	\caption{\label{fig:PrefPastProperTimes} Probability distributions for the proper time separations between each causet event and its preferred past, according to the 6 criteria. The histograms have a bin size of a twentieth of the time scale $\sqrt[d]{1 / \rho}$ and are observed for the region $U_{2}$ in sprinkles with an expected cardinality of 6000.}
\end{figure}
We compare the statistics of proper time separation (measured in length units $\sqrt[d]{1 / \rho}$) between an event and its rank 2 past events selected by each criterion, see \autoref{fig:PrefPastProperTimes}. 
Once again, we display the proper time distributions in dimensions $d = 1 + 1$ (top), $d = 1 + 2$ (middle) and $d = 1 + 3$ (bottom), using the observation region $U_{2}$ (see \eqref{eq:VolReductionFactors}). 

Criteria 4 and 5 yield proper time distributions that broaden with increasing spacetime dimension, while the peaks of criteria 2, 3, and 6 are more pronounced and get sharper with increasing spacetime dimension. 
In $1 + 3$ dimensional Minkowski space, about 70\% of the subsets selected by criterion 3 (largest pure diamonds) are singleton (see \autoref{fig:PrefPasts}) so that the same subsets are selected by criterion 6 as well. 
Other singletons selected by criterion 6 span diamonds with almost the same size. 
This is reflected in very similar proper time distributions for criteria 3 and 6 in dimension $1 + 3$. 

Criterion 1 yields the worst result here, since the diamonds corresponding to the rank 2 past events in $D^{-}_{\mathrm{crit\,1}}( x )$ (for a causet event $x \in \mathscr{C} \setminus C^{-}_{2}$) can have any size almost up to the entire past of $x$ in $\mathscr{C}$.  
In \autoref{fig:PrefPastProperTimes}, the probability densities for criterion 1 reach their maxima at approximately 37 for dimension $d = 1 + 1$, at 14.5 for dimension $d = 1 + 2$ and around 7.2 for dimension $d = 1 + 3$ in  units $\sqrt[d]{1 / \rho}$, thus falling far beyond the plotting range of the proper time axes. 

When looking at the proper time separation, we find that criterion 6, followed by criterion 3 and 2 perform best giving a probability distribution with relatively low expectation value and variance. 

\subsection{Distribution of the rank 2 past subsets along the unit hyperboloid}
\begin{figure}
	\centering
	\includegraphics[scale=\graphicscaling]{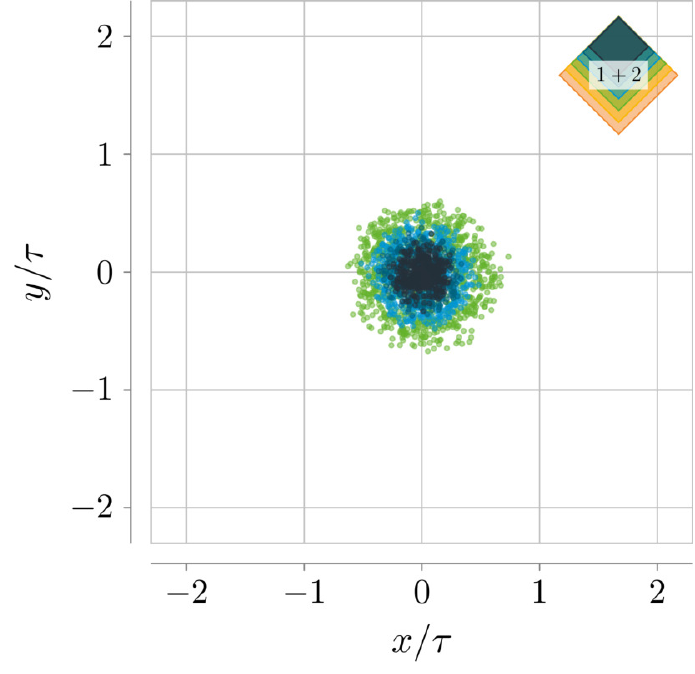}
	\includegraphics[scale=\graphicscaling]{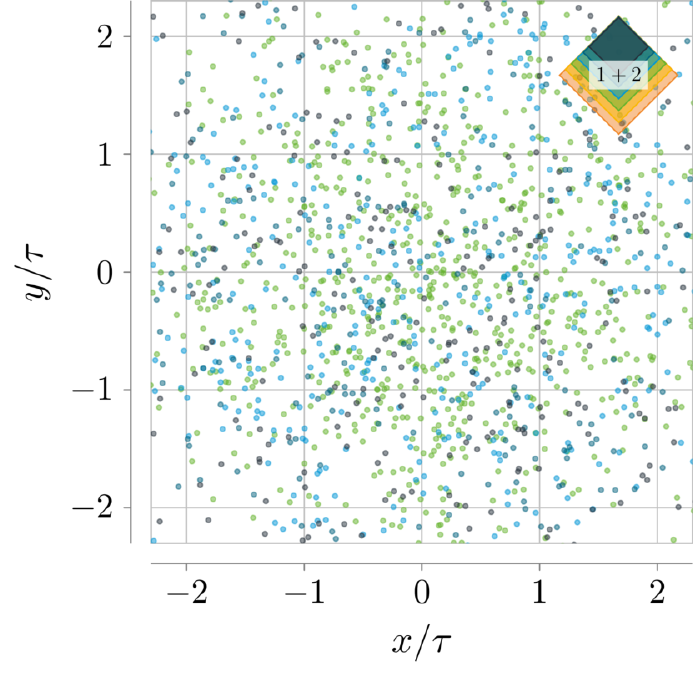}
	\caption{\label{fig:PrefPastHyperboloidsScatter} Scatter plots of the rank 2 past events distributed along the unit hyperboloid for dimensions $1 + 2$ and all observation regions $U_{i}$ ($i \in \{ 2, 3, 4, 5 \}$ from lighter to darker shades, green, blue, dark blue, black) for criteria 1 (left) and 6 (right). Both plots are for single sprinkles with about 6000 events.}
\end{figure}

\begin{figure}
	\centering
	\includegraphics[scale=\graphicscaling]{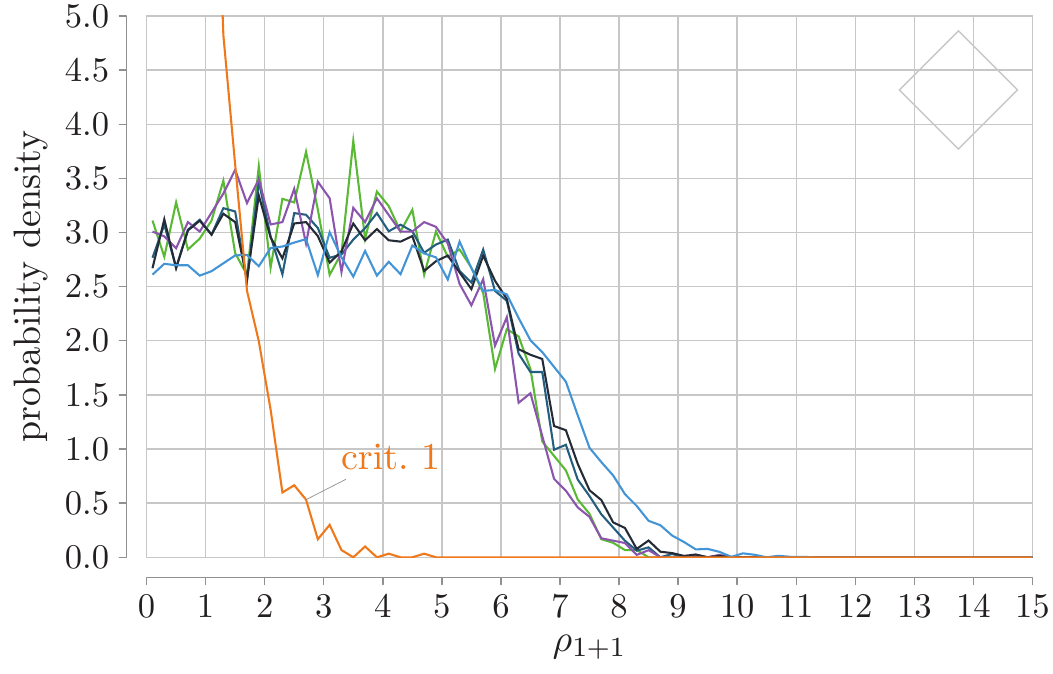}
	\includegraphics[scale=\graphicscaling]{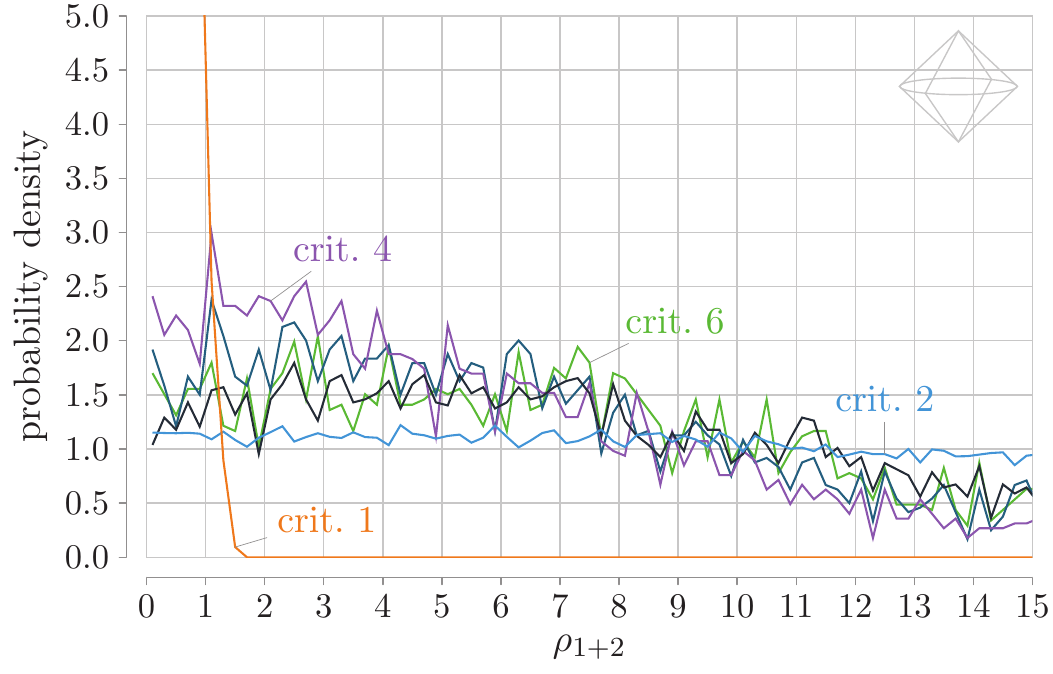}
	\includegraphics[scale=\graphicscaling]{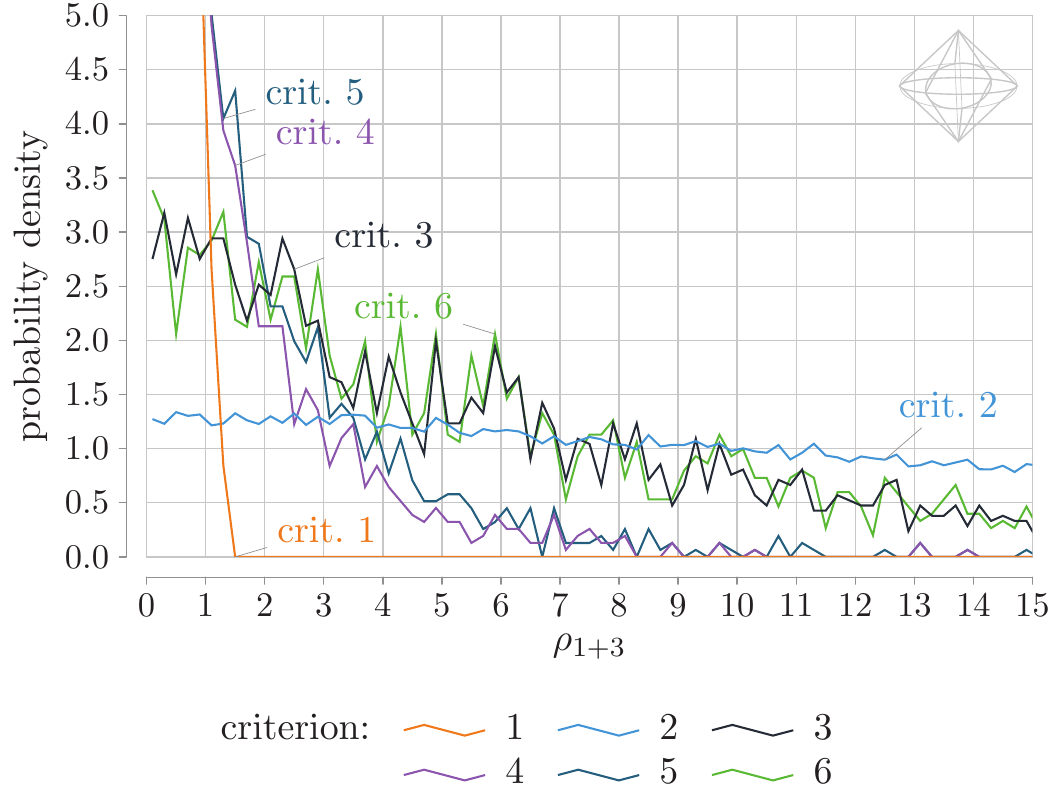}
	\caption{\label{fig:PrefPastHyperboloids} Probability distribution of the rank 2 past subsets projected onto the unit hyperboloid, for all 6 criteria at observation volume $U_{2}$ for a single sprinkle (of about 6000 events) in Minkowski spacetime of dimension $1 + 1$ (top), $1 + 2$ (middle) and $1 + 3$ (bottom).}
\end{figure}

Even though discrete subsets of Minkowski spacetime like a sprinkle break Lorentz symmetry, the entire configuration space for the spacetime (see details in Sec.~\ref{sec:InfiniteSprinkles}) is Lorentz invariant since it includes all transformed versions of the sprinkle. 
If the distributions of rank 2 past events selected by most criteria are uniform in the limit of large sprinkles, the selected subsets tend to be Lorentz invariant. 
We check this by viewing the relative coordinates $( x_{0}, x_{1}, \dotso )$ of all rank 2 past events $D^{-}_{\mathrm{crit}\,n}( x )$ corresponding to the criterion $n$ with respect to event $x$ and project it onto the unit past hyperboloid, i.e.\ dividing by the proper time separation 
\begin{align}
	\label{eq:ProperTimeSeparation}
			\tau 
	&= \sqrt{x^{2}_{0} - r^2} 
	, & \text{where}\qquad 
			r^2 
	&= \sum_{i = 1}^{d - 1} x_{i}^2 
	. 
\end{align}
For example, see the scatter plots for crtieria 1 and 6 at dimension $1 + 2$ in \autoref{fig:PrefPastHyperboloidsScatter}. 
The data points in the scatter plots are shaded corresponding to the observation region $U_{i}$ ($i \in \{ 2, 3, 4, 5 \}$) to which the event $x$ belongs. 
These plots suggest, by eye, that the events selected by criterion 1 tend to cluster on the unit hyperboloid while those selected by criterion 6 are more uniformly distributed. 
This is investigated more systematically in the following and shown in \autoref{fig:PrefPastHyperboloids}.

In order to obtain the graphs of \autoref{fig:PrefPastHyperboloids}, we define the rescaled coordinate $\rho_{d}$ such that its differential $\d\rho_{d}$ describes equal volume slices along the radial direction of the unit hyperboloid. 
It is the product of the volume of the unit $d - 2$ sphere $S_{d - 2}$ and the integral over the hyperbolic radius up to the value $u' = \arsinh( r / \tau )$, 
\begin{align}
	\label{eq:HyperboloidPlotScaling}
			\rho_{d} 
	&= \int_{S_{d - 2}} \d{\varOmega_{d - 2}} 
			\int_{0}^{u'} \sinh^{d - 2}( u ) \id{u} 
	. 
\end{align}
Evaluate these integrals to find $\rho_{d}$ as function of the normalized radial coordinate $r / \tau$, 
\begin{subequations}
\begin{align}
	\label{eq:HyperboloidPlotScalingD2}
			\rho_{1 + 1} 
	&= 2 \arsinh \frac{r}{\tau} 
	, 
	\\\label{eq:HyperboloidPlotScalingD3}
			\rho_{1 + 2} 
	&= 2 \pi \left[ \sqrt{1 + \left( \frac{r}{\tau} \right)^2} - 1 \right] 
	, 
	\\\label{eq:HyperboloidPlotScalingD4}
			\rho_{1 + 3} 
	&= 4 \pi \left[ 
				\frac{1}{4} 
				\sinh\left( 2 \arsinh \frac{r}{\tau} \right) 
			- \frac{1}{2} 
				\arsinh \frac{r}{\tau} 
			\right]  
	. 
\end{align}
\end{subequations}
Note that the hyperbolic radius $u' = \arsinh( r / \tau )$ is the rapidity (with respect to the inertial coordinates) of the inertial motion connecting the point on the hyperboloid to the origin. 

All criteria but the first yield a constant distribution falling off at values close to the outer boundary of the sprinkling region at dimensions $1 + 1$ and $1 + 2$. 
Boundary effects at dimension $1 + 3$ are more pronounced, so that we only have criteria 2, 3, and 6 with a close to uniform distribution. 
Criterion 1 has a strong bias to select rank 2 events close to the origin for all investigated Minkowski spacetimes, because the selected events correspond to the largest diamond so that it tends to be as close as possible to the bottom tip of the entire sprinkling region $U$. 
Comparing similar plots for all observation regions $U_{i}$ (not shown), we find that the distributions are getting more and more homogeneous from $i = 0$ to $i = 5$, except for criterion 1, which concentrates more and more around the origin when decreasing the observation region. 

In combination of the characteristics, we find that criterion 6 has the highest probability for a unique rank 2 past event, while also yielding a sharply peaked proper time distribution with a maximum at low proper time separation, and tend to give a uniform distribution along the unit hyperboloid. 
The next part of this section is focused on this criterion only. 

\subsection{Diamond sizes for criterion 6}
\label{sec:Rank2PastCriterion6DiamondSizes}
The analysis presented in the previous part of the section indicates that criterion 6 has the best performance among those studied. 
We proceed to investigate criterion 6 in more depth by computing the resulting distribution of diamond sizes spanned between the causet event $x$ and the elements of $D^{-}_{\mathrm{crit\,6}}( x )$. 
\begin{figure*}[p]
	\centering
	\includegraphics[scale=\graphicscaling]{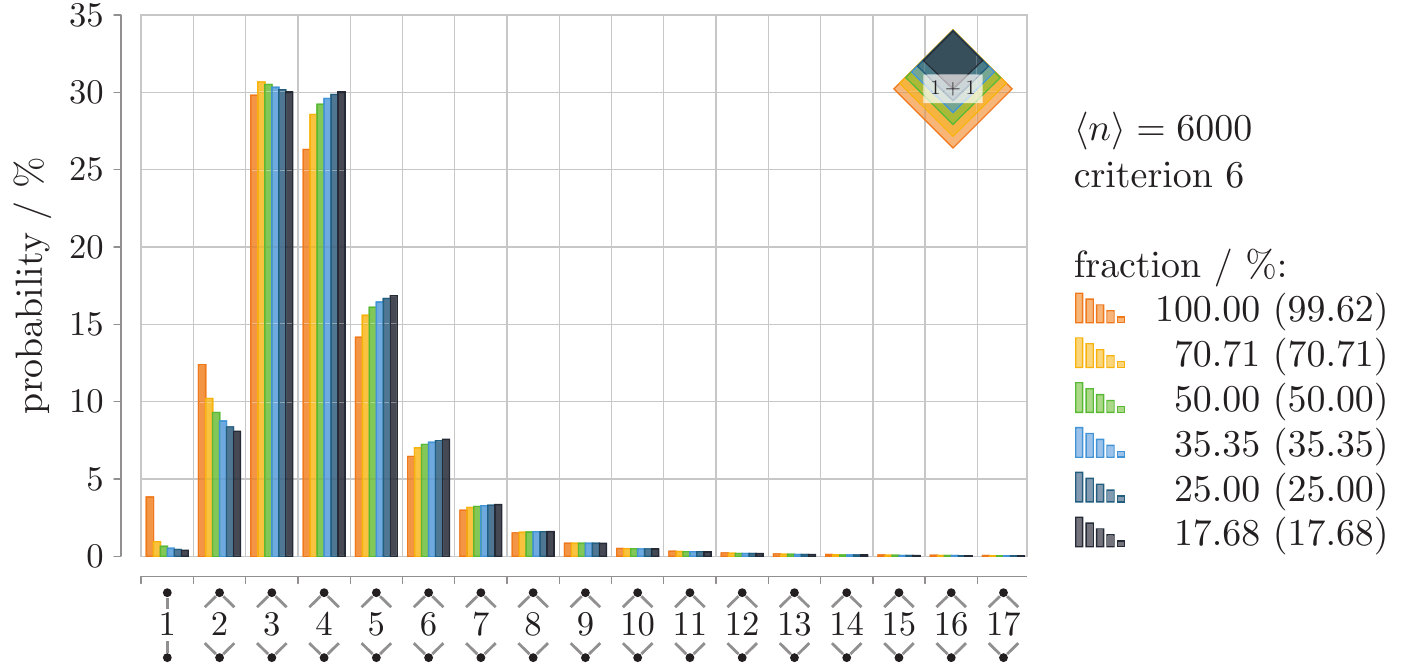}
	\includegraphics[scale=\graphicscaling]{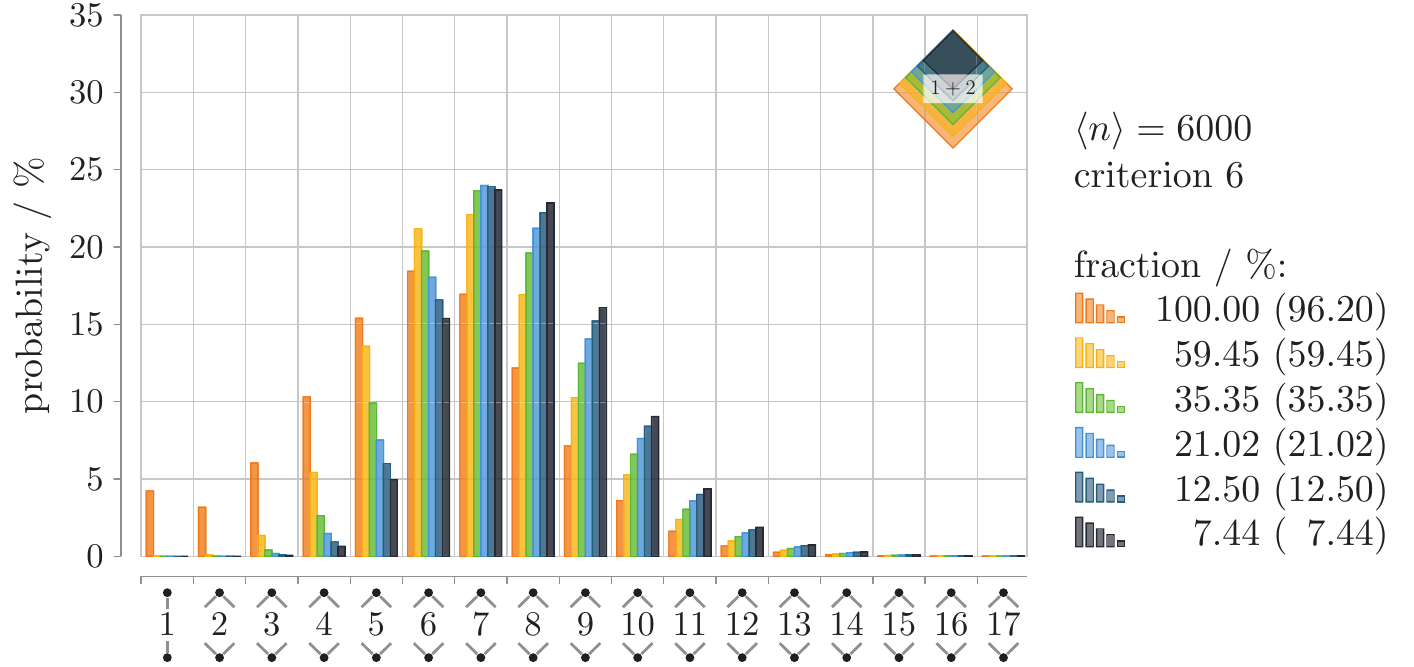}
	\includegraphics[scale=\graphicscaling]{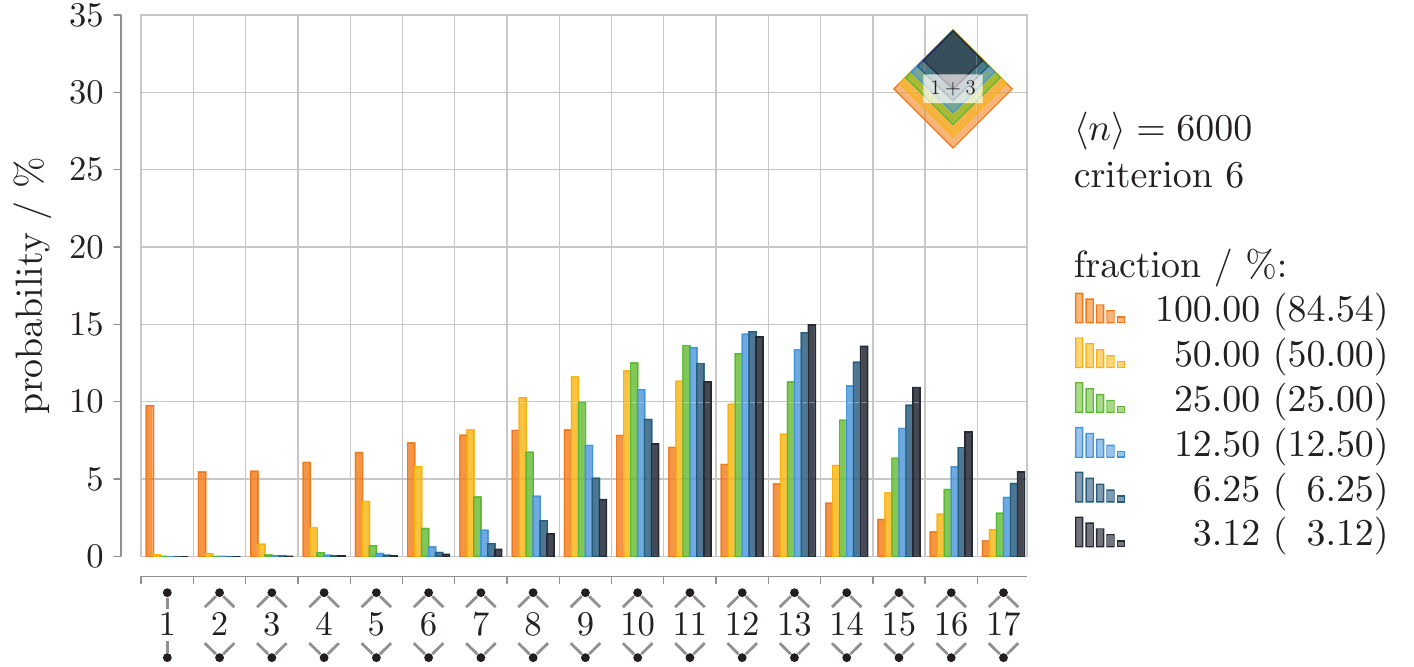}
	\caption{\label{fig:DiamondsPref5} Discrete probability distribution of diamond cardinalities and how they change by shrinking the observed region $U_{i}$ from $i = 0$ (light/orange shade) to $i = 5$ (black shade). The plot legends show the observation volume (and causet fraction that has at least one preferred past). The spacetime dimension increases from $1 + 1$ (top plot) to $1 + 3$ (bottom plot). Along the horizontal axis, the histogram bins are labelled by the diamond size.}
\end{figure*}
The plots in \autoref{fig:DiamondsPref5} show the probability distribution of the diamond size. 
The plot legends list the actual size of the respective observed volumes as fraction of the entire sprinkling region. 
In brackets, we denote the fractions of the causets that are within the observation regions and have a non-empty rank 2 past. 

Note the change of the histograms when reducing the observation region $U_{i}$ from $i = 0$ (light/orange shade) to $i = 5$ (black shade), because the diamonds that are getting smaller towards the past infinity are excluded. 
Especially for $d = 1 + 3$, the reduction from observation region $U_{0}$ to the first smaller region $U_{1}$ causes a strong increase in the diamond sizes. 

\subsection{Diamonds along a timelike geodesic path}
\label{sec:GeodesicPath}
The causal set analogue of a timelike geodesic is defined as follows.
\begin{definition}
	A \defof{(timelike) geodesic} between 2 events $x \preceq y \in \mathscr{C}$ of a causet $\mathscr{C}$ is a path in $\paths( x, y )$ with maximal cardinality~\cite{1978Myrheim}. 
\end{definition}
For all sprinkles in our simulations, we also investigate the diamonds that are spanned by next-to-nearest neighbours along the timelike geodesic paths between the events with minimum and maximum time coordinate. 
See \autoref{fig:VolReductionGeodesicPath} above for an example. 
The paths have an expected length of 137.4 events in $1 + 1$, 26.2 events in $1 + 2$, and 8.1 events in $1 + 3$ dimensional Minkowski spacetime, again taking the average over 10000 sprinkles for each dimension. 

\begin{figure}
	\centering
	\includegraphics[scale=\graphicscaling]{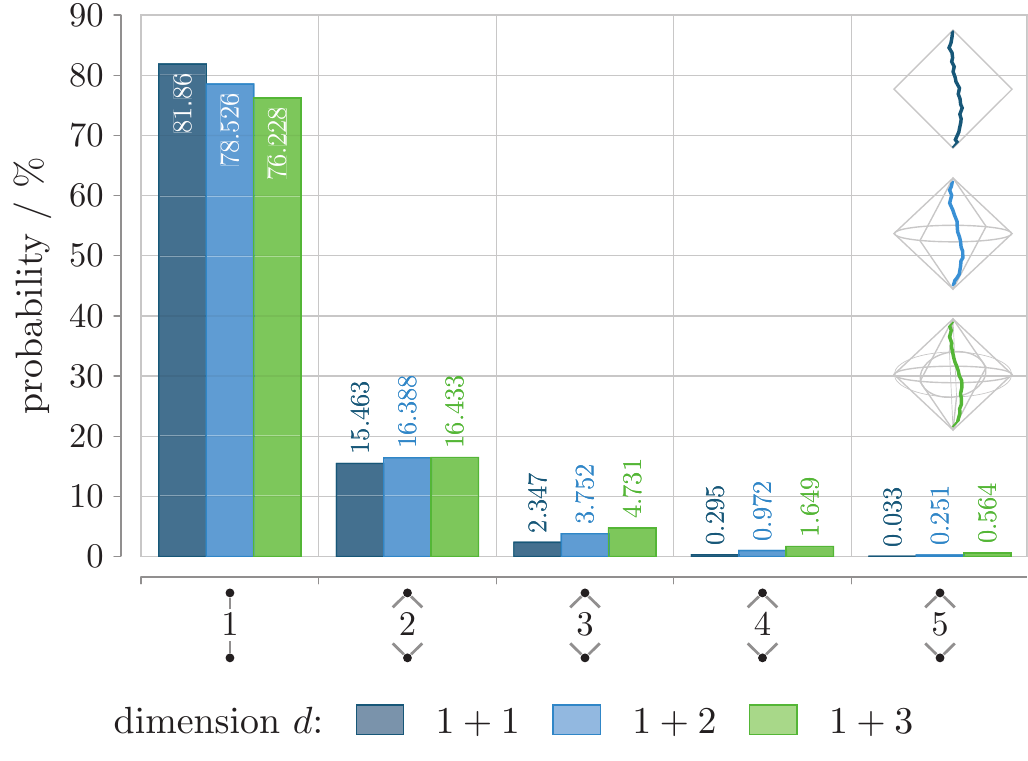}
	\includegraphics[scale=\graphicscaling]{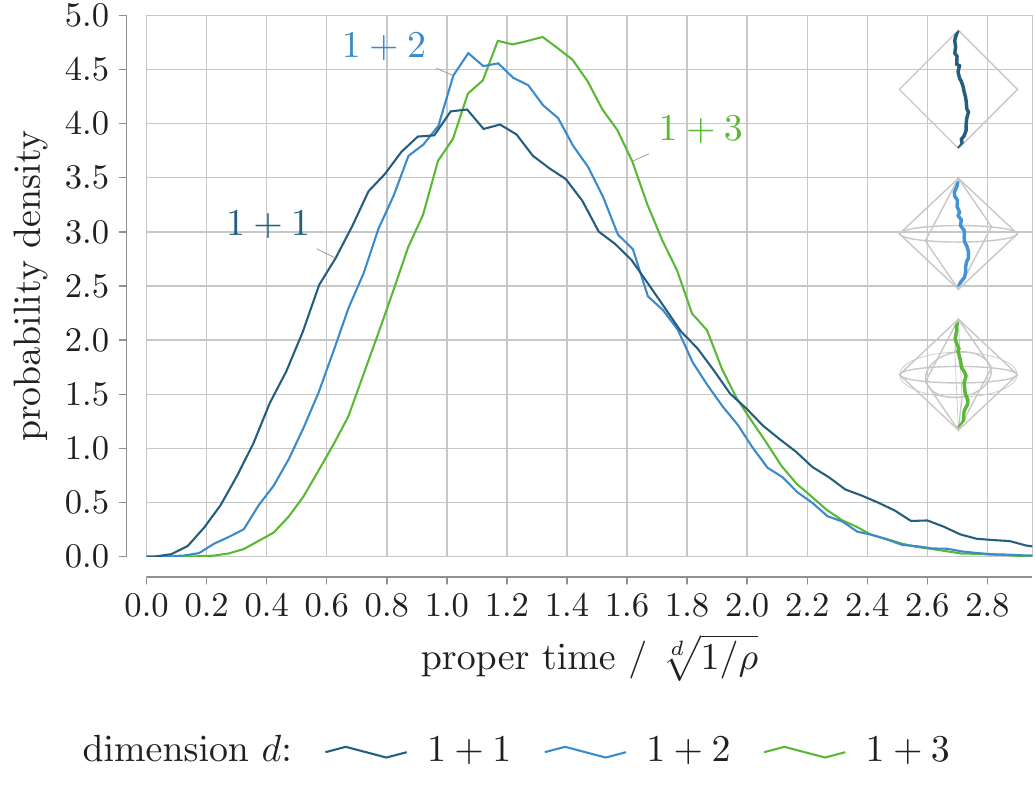}
	\caption{\label{fig:Geodesic} Probability distributions of diamond sizes (top plot) and their proper time separations (bottom plot) for the diamonds along all timelike geodesics from the bottom to the top through the sprinkles (with an expected cardinality of 6000).}
\end{figure}
In \autoref{fig:Geodesic}, we show the probability distributions for the size of the diamonds and the proper time separation spanned by them along the geodesics. 
These statistics are taken for the entire sprinkling region (without reducing the observation region), since the diamonds along the geodesic are not influenced by the past or future infinities of the finite sprinkles. 

The histograms in the top plot of \autoref{fig:Geodesic} peak at 1-diamonds and barely differ for the three Alexandrov subsets of flat spacetimes in dimensions $1 + 1$, $1 + 2$ and $1 + 3$. 
An observer travelling along a geodesic path in a causet might be unable to identify the spacetime dimension from the local structure of diamonds. 
Notice that each diamond along this path can only be a pure diamond, since every geodesic is a maximal path. 

Because each sprinkle is embedded, we can assign the proper time separation to the diamonds and thus determine a ticking rate of a `diamond clock' for an observer following a geodesic path. 
The results are shown in the second plot of \autoref{fig:Geodesic}. 
The proper time statistics have very similar expectation values: 
\begin{subequations}
\begin{align}
	\label{eq:ProperTimesChainsExpectation2D}
			\langle \tau \rangle_{1 + 1} 
	&= 1.236 \rho^{-1/2} 
	, 
	\\\label{eq:ProperTimesChainsExpectation3D}
			\langle \tau \rangle_{1 + 2} 
	&= 1.193 \rho^{-1/3} 
	, 
	\\\label{eq:ProperTimesChainsExpectation4D}
			\langle \tau \rangle_{1 + 3} 
	&= 1.278 \rho^{-1/4} 
	. 
\end{align}
\end{subequations}
In sprinkling units (such that $\rho = 1$), the expected proper times are quite close and a clock that is ticking in accordance with the diamonds shows a similar time passing along an equal path length (number of diamonds) in all dimensions. 
The distributions of the proper time separation are also very similar to the results for criterion 2 in dimension $1 + 1$, see \autoref{fig:PrefPastProperTimes}, since that criterion selects the smallest diamonds. 
However, here we observe almost the same statistics for the timelike geodesic paths in sprinkling dimensions $1 + 2$ and $1 + 3$. 
This evidence supports the discussion from~\cite{2015Carlip} suggesting a dimensional reduction to dimension $1 + 1$ for an observer in causal set theory at very small scales. 
For a conclusive argument, however, further investigations including dimension estimators (like the Myrheim-Meyer estimator~\cite{1978Myrheim,1988Meyer}) are necessary. 

\section{(Infinite) sprinkled causal sets on a spacetime manifold}
\label{sec:InfiniteSprinkles}
We construct the probability space for sprinkling on any globally hyperbolic spacetime.
In this paper, we will use the Poisson probability measure primarily to compute the expectation values of past infinity cardinalities in sprinkled causets. 
However, the precise construction of the probability space may also find other applications. 
For example, it may facilitate a more general discussion of the continuum limit of causal sets, like previously considered for compact spacetime manifolds in \cite{2000Bombelli}. 

Throughout this section, let $M$ be a fixed globally hyperbolic, $d$-dimensional spacetime manifold with metric $g$. 
The Poisson process called sprinkling randomly selects a finite subset as a causal set from the spacetime $M$~\cite{2009Sorkin}. 
For the construction of a probability space for the sprinkling process, we need to find a space of ``possible outcomes'', an appropriate class of measurable subsets (a $\upsigma$-algebra), and a probability measure. 

For the sprinkling process, the sample space -- or configuration space -- is given by the set of all discrete subsets of the manifold $M$. 
We review the configuration space for $M$ and compact subsets of the manifold $U \subset M$ as introduced by~\cite{1998AlbeverioKondratievRoeckner}. 
This framework was constructed for applications in quantum theory~\cite{1999AlbeverioKondratievRoeckner} but, as we will show, may be applied to causal set theory. 
The construction provides a Borel $\upsigma$-algebra over the configuration space and leads to the discussion of the Poisson probability measure for the subsets $U$ and the entire manifold $M$. 

For examples of sprinkling in an Alexandrov subset of $1 + 1$ dimensional Minkowski space, we compute the probability for sprinkling a given causet by counting 2D-orders. 
This leads to the computation of the probability for a causet event to be in the 1- and 2- layer past infinity and their asymptotic behaviour for infinite sprinkles on $1 + 1$ dimensional Minkowski space. 

\subsection{The sprinkling probability space}
In the literature, the term `sprinkling' is used to refer to the random Poisson process as well as to an element of the configuration space~\cite{2009Sorkin}. 
Here we want to make the notions more distinct and put the sprinkling process in a more formal language. 
\begin{definition}
	The \defof{sprinkling configuration space} is the set of all locally finite subsets of $M$, 
	\begin{align}
		\nonumber
				Q 
		&:= \bigl\{ 
					S \subset M 
				\bigm| 
					\forall \text{ compact } U \subset M: 
		\\\label{eq:SprinklingConfig}
		&\qquad\qquad\qquad
					| S \cap U | < \infty 
				\bigr\}
		. 
	\end{align}
	We call each element of this configuration space a \defof{sprinkle} on $M$ and refer to the Poisson process of constructing causets for $M$ as \defof{sprinkling}. 
\end{definition}
	
To find the $\upsigma$-algebra $\Borelset( Q )$ over this configuration space -- the space of subsets of $Q$ to which we can assign a probability -- first, consider a compact subset $U \subset M$. 
The sprinkling configuration space $Q_{U}$ for $U$ is the (disjoint) union of configuration spaces $Q_{U, n}$ with fixed cardinalities $n$, 
\begin{align}
	\label{eq:CSSprinklingConfigN}
			Q_{U, n} 
	&:= \left\{ 
				S \subset U 
			\bypred 
				| S | = n 
			\right\} 
	, 
	\\\label{eq:CSSprinklingConfig}
			Q_{U} 
	&= \bigcup_{n = 0}^{\infty} Q_{U, n} 
	. 
\end{align}
The $n$-fold Cartesian product $U^{n}$ has the $n$-fold product topology.  
Let $F_{n}$ denote the fat diagonal of the Cartesian product, which is the subset of all $n$-tuples that have at least one pair of identical components. 
Deleting $F_{n}$, we obtain the configuration space of $n$  indistinguishable points in $U$, 
\begin{align}
	\label{eq:CSSprinklingTubleConfig}
			\tilde{Q}_{U, n} 
	&:= U^{n} \setminus F_{n} 
	, 
\end{align}
with the subspace topology. 
As there is no physical significance to the order in which a set of spacetime events is listed, the configuration space $Q_{U, n}$ is the image of 
\begin{align}
	\nonumber
			\varSigma_{U, n} : \tilde{Q}_{U, n} 
	&\to Q_{U, n} 
	, \\\label{eq:CSSprinklingTubleConfigLabelling}
			( x_{1}, x_{2}, \dotso, x_{n} ) 
	&\mapsto \{ x_{1}, x_{2}, \dotso, x_{n} \} 
	, 
\end{align}
which maps all $n!$ permutations of some $n$-tuple to the same set of $n$ events. 
So the configuration space $Q_{U, n}$ is the quotient by the $n$-th symmetric group acting on the $n$-tuples and endowed with the quotient topology induced by $\varSigma_{U, n}$. 
The elements of the Borel $\upsigma$-algebra $\Borelset( Q_{U, n} )$ are generated by subsets that are open on $\tilde{Q}_{U, n}$ under the pre-image of $\varSigma_{U, n}$. 
Take the disjoint union over $n$, see \eqref{eq:CSSprinklingConfig}, to find the disjoint union topology on $Q_{U}$ and the Borel $\upsigma$-algebra $\Borelset( Q_{U} )$. 
Finally, the inverse limit over all configuration spaces $Q_{U}$ leads to the Borel $\upsigma$-algebra $\Borelset( Q )$. 

So far, we have the configuration space $Q$ with the Borel $\upsigma$-algebra. 
For the probability space $( Q, \Borelset( Q ), \mu )$, it remains to specify the Poisson probability measure 
\begin{align}
	\label{eq:SprinklingProbabilityMeasure}
			\mu : \Borelset( Q ) 
	&\to [ 0, 1 ] 
\end{align}
that corresponds to the sprinkling process on $M$ with a given positive constant that we call the sprinkling density $\rho$. 
The measure restricted to any compact subset $U \subset M$ is derived from the metric induced volume measure on the spacetime $( M, g )$. 
For every measurable spacetime subset $O \in \Borelset( M )$ the volume measure $\nu$ assigns a positive real value 
\begin{align}
	\label{eq:VolumeMeasure}
			\nu( O ) 
	&:= \int_{O} \sqrt{| g |} \id[d]{x} 
\end{align}
(including the metric factor in some coordinate chart). 
The sprinkling measure is defined using the product measure $\nu^{n}$ for subsets of the Cartesian product $U^{n}$ and its push-forward by the map $\varSigma_{U, n}$. 
\begin{definition}
	The \defof{Poisson (probability) measure $\mu_{U}$} with \defof{sprinkling density $\rho$ on any compact subset $U$} of the manifold $M$ with volume measure $\nu$ assigns a probability to each subset in $\Borelset( Q_{U} )$ such that for every $n \in \fieldN[0]$ and $B_{n} \in \Borelset( Q_{U, n} )$ 
	\begin{align}
		\label{eq:CSSprinklingProbMeasure}
				\mu_{U}( B_{n} ) 
		&= \e^{- \rho \nu( U ) } 
				\frac{\rho^{n}}{n!} 
				\nu^{n} \left( \varSigma_{U, n}^{-1}( B_{n} ) \right) 
		. 
	\end{align}
	where $\varSigma_{U, n}^{-1}$ denotes the pre-image by \eqref{eq:CSSprinklingTubleConfigLabelling}. 
\end{definition}
Note that the exponential factor normalizes $\mu_{U}$ to a probability measure so that $\mu_{U}( Q_{U} ) = 1$. 
The sprinkling process on $M$ is obtained by an inverse limit over the measures
$\mu_{U}$ for all compact subsets $U \subset M$ as an application of the Hahn-Kolmogorov theorem so that the Poisson measure $\mu$ is uniquely determined by the measure family of $\mu_{U}$ for all $U$~\cite[Theorem 4.2]{2005Parthasarathy}. 
It is described by a measure $\mu$ on $Q$ with the following property: for all compact subsets $U \subset M$ and all $B \in \Borelset( Q_{U} )$, 
\begin{align}
	\nonumber
			\mu( \hat{B}_{U} ) 
	&= \mu_{U}( B ) 
	, \\\label{eq:SprinklingProbMeasureProperty}
			\text{where}\qquad 
			\hat{B}_{U} 
	&:= \left\{ 
				S \in Q 
			\bypred 
				S \cap U \in B 
			\right\}
	. 
\end{align}
It is shown in~\cite[sec.\ 2.2]{1998AlbeverioKondratievRoeckner} that the following integral formula holds 
\begin{align}
	\nonumber
	&
			\int_{Q} 
				\exp\Biggl( \sum_{x \in S} f( x ) \Biggr) 
			\id{\mu}( S ) 
	\\\label{eq:SprinklingProbMeasureDefinition}
	&= \exp\Biggl( 
				\rho 
				\int_{M} 
					\Bigl( \e^{f( x )} - 1 \Bigr) 
				\id{\nu}( x ) 
			\Biggr) 
\end{align}
for every compactly supported, continuous function $f$ on $M$, where the integral on the left hand side runs over all sprinkles $S \in Q$.  
This provides an alternative definition of $\mu$; see~\cite[sec.\ 2]{1998AlbeverioKondratievRoeckner} for more details. 

In summary, this construction yields the probability spaces $( Q_{U}, \Borelset( Q_{U} ), \mu_{U} )$ for all compact subsets $U \subset M$ and $( Q, \Borelset( Q ), \mu )$ for the manifold $M$. 

\subsection{Causet isomorphism classes}
Given any sprinkle $S \in Q$, the partial order $x \preceq y$ for $x, y \in S$ is the causal relation of $M$ restricted to the subset $S$. 
It is given by the future and past $J^{\pm}$ such that 
\begin{align}
	\label{eq:CSSprinklingCausality}
			x \preceq y 
	&\liff x \in J^{-}( y ) 
	, 
\end{align}
where $J^{-}( y ) \subset M$ is the set of all events that are in the causal past of $y$. 
Two sprinkles are isomorphic, denoted by the symbol $\sim$, if there exists a bijection between them that preserves the causal relation. 
The sprinkles $S \in Q$ that are isomorphic to a given causet $\mathscr{C}$ form an isomorphism class $[ \mathscr{C} ] \subset Q$, which is the set of all possible embeddings of the causet $\mathscr{C}$ in $M$. 

For every compact subset $U \subset M$, all sprinkles $S \in Q_{U}$ have a finite cardinality $| S | = n$ and, for any fixed cardinality $n$, there is a finite number $a( n )$ of distinct causet isomorphism classes, forming an integer sequence labelled as A000112~\cite{website:oeisA000112}. 
A closed expression for the term $a( n )$ of this sequence is unknown, but the first 16 terms have been computed~\cite{2002BrinkmannMcKay} -- and the first 12 terms are given in \autoref{tab:OEIS.A000112}. 
To the right of the table, we show all $a( n ) = 16$ causets for $n = 4$ as Hasse diagrams. 
\begin{table*}
	\centering
	\includegraphics[scale=\graphicscaling]{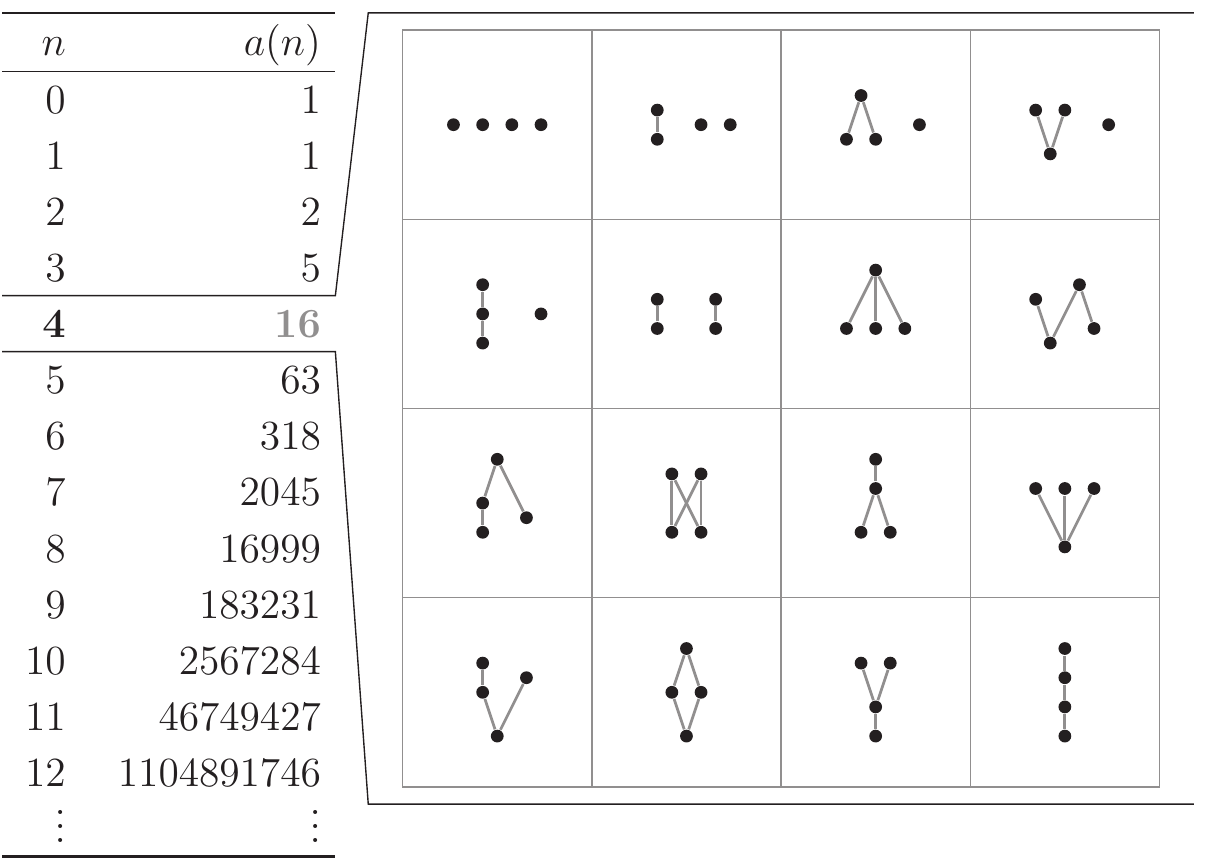}
	\caption{\label{tab:OEIS.A000112} First elements of the integer sequence A000112: number of partial ordered sets (posets) with $n$ indistinguishable elements. The Hasse diagrams for all possible 4-event causets are shown on the right.}
\end{table*}
For cardinalities up to 5, one can embed every causet in an Alexandrov subset of $1 + 1$ dimensional Minkowski spacetime. 
For cardinalities greater than 5, there exist causets that cannot be embedded in $1 + 1$ dimensional flat spacetime. 

We can use causet isomorphism classes $[ \mathscr{C} ] \subset Q_{U}$ to compute the probability that a random sprinkle $\mathsf{S}$ (arising from sprinkling into $U$ with density $\rho$) has the causal relations of $( \mathscr{C}, \preceq )$, 
\begin{align}
	\label{eq:RandomVariableProbability}
			\Pr( \mathsf{S} \sim \mathscr{C} ) 
	&= \mu_{U}\bigl( [ \mathscr{C} ] \bigr) 
	. 
\end{align}
Here we use the fact that any causet isomorphism class $[ \mathscr{C} ]$ is a measurable set in $\Borelset( Q_{U} )$ (ultimately due to the causal relation describing a closed subset of $Q_{U}$). 
For example, the causet $\mathscr{C}_{12}$ of 2 causally related events has equivalence class
\begin{align}
	\label{eq:CSSprinkleIsomorphismClass}
			[ \mathscr{C}_{12} ] 
	&= \bigl\{ 
				\{ x_1, x_2 \} \in Q_{U} 
			\bigm| 
				x_1 \prec x_2 
			\bigr\} 
	,
\end{align}
with pre-image
\begin{align}
	\nonumber
			\varSigma_{U, 2}^{-1}\bigl( [ \mathscr{C}_{12} ] \bigr) 
	&= \bigl\{ 
				\{ x_1, x_2 \} \in \tilde{Q}_{U, 2} 
			\bigm|
	\\\label{eq:CSSprinkleIsomorphismClassPreimage}
	&\qquad\qquad 
				(x_1 \prec x_2)\vee (x_2 \prec x_1)
			\bigr\} 
\end{align}
in $\tilde{Q}_{U, 2}$. 
Using \eqref{eq:CSSprinklingProbMeasure}, and noting that $\varSigma_{U, 2}^{-1}([\mathscr{C}_{12}])$ consists of $2! = 2$ disjoint sets of equal volume,
it follows that the probability of
sprinkling a causal set of this type into $U$ is
\begin{align}
	\nonumber
			\Pr\left( \mathsf{S} \sim \mathscr{C}_{12} \right) 
	&= \mu_{U}\bigl(	[ \mathscr{C}_{12} ] \bigr) 
	\\\nonumber
	&= \e^{- \rho \nu( U )} \rho^2 
	\\\label{eq:CSSprinkleTwoCausallyRelatedEvents}
	&\quad{}\times
			\int_{U} \d{\nu}( x_1 ) 
			\int_{J^{+}( x_1 ) \cap U} \d{\nu}( x_2 ) 
	. 
\end{align}

\subsection{2D-orders and the past infinity of causets}
\label{sec:2DOrdersAndPastInf}
In the following, we use the correspondence between sprinkles on an Alexandrov subset $U$ of $1 + 1$ dimensional Minkowski spacetime and their \emph{2D-orders}~\cite{1985Winkler,1990Winkler} to compute probabilities of obtaining an element of any given causet isomorphism class when sprinkling into $U$. 
A 2D-order is the product of two total orders. 

\begin{figure}
	\centering
	\includegraphics[scale=\graphicscaling]{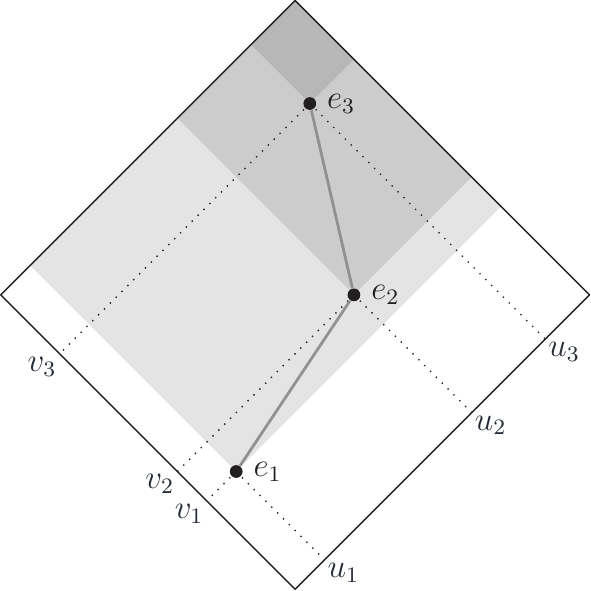}
	\hspace{2em}
	\caption{\label{fig:PosetProbN3D2bicone123} Sprinkle with three events in an Alexandrov subset of $1 + 1$-dimensional Minkowski spacetime that is isomorphic to the 3-chain causet $\mathscr{C}_{123}$. The futures of the events are shaded.}
\end{figure}

Starting with an example, consider the sprinkles $S \in Q_{U}$ that are isomorphic to the chain causet $\mathscr{C}_{123}$ of 3 events. 
Using standard null coordinates $( u, v )$ and excluding a set of measure zero, we restrict to those sprinkles comprising events at $( u_i, v_i ), i \in \{ 1, 2, 3 \}$ such that $u_1 < u_2 < u_3$ and $v_1 < v_2 < v_3$. 
One such sprinkle is pictured in \autoref{fig:PosetProbN3D2bicone123}. 
Using \eqref{eq:CSSprinklingProbMeasure}, the probability for a random sprinkle $\mathsf{S}$ into $U$ to be isomorphic to $\mathscr{C}_{123}$ is 
\begin{align}
	\nonumber
			\Pr( \mathsf{S} \sim \mathscr{C}_{123} ) 
	&= \left( 3! 
				\int_{0}^{1} \d{u_1} 
				\int_{u_1}^{1} \d{u_2} 
				\int_{u_2}^{1} \d{u_3} 
			\right.
	\\\nonumber
	&\quad\qquad{}\times
			\left.
				\int_{0}^{1} \d{v_1} 
				\int_{v_1}^{1} \d{v_2} 
				\int_{v_2}^{1} \d{v_3} 
			\right) 
	\\\label{eq:PosetProbN3D2bicone123}
	&\quad{}\times
			\left( \e^{- \rho \nu( U )} \frac{\rho^3}{3!} \nu( U )^3 \right)
\end{align}
\begin{align}
	\label{eq:PosetProbN3D2bicone123Result}
			\Pr( \mathsf{S} \sim \mathscr{C}_{123} ) 
	&= \frac{1}{3!} 
			\left( \e^{- \rho \nu( U )} \frac{\rho^3}{3!} \nu( U )^3 \right) 
	, 
\end{align}
where we have pulled out a volume factor and correspondingly scaled the null coordinates such that they range over the unit interval. 
The 6-fold integral in \eqref{eq:PosetProbN3D2bicone123} has a factor of 3!, since there are that many distinct labellings (total orders) of the events by their $u$-coordinate. 
The bracketed expression in \eqref{eq:PosetProbN3D2bicone123Result} is the probability for a random sprinkle to have 3 events, 
\begin{align}
	\nonumber
			\Pr\bigl( | \mathsf{S} | = 3 \bigr) 
	&= \mu_{U}( Q_{U, 3} ) 
	\\\label{eq:D2BiconeCausetProbN3}
	&= \e^{- \rho \nu( U )} \frac{\rho^3}{3!} \nu( U )^3 
	. 
\end{align}
Thus by Bayes' theorem, the remaining factor must be the conditional probability for ``the sprinkled causet is the 3-chain given that the sprinkle has 3 events'', 
\begin{align}
	\label{eq:D2BiconeCausetProb123WhenN3}
			\Pr\bigl( \mathsf{S} \sim \mathscr{C}_{123} \bigm| | \mathsf{S} | = 3 \bigr) 
	&= \frac{1}{3!} 
	. 
\end{align}
Now, we consider another method to determine this conditional probability by combinatorial means. 

Let $S$ be any finite sprinkle on an Alexandrov subsets of $1 + 1$ Minkowski spacetime, comprising events with null coordinates $( u_{i}, v_{i} )$ ($1 \leq i \leq n = | S |$). 
We will say that $S$ is non-degenerate if all the $u$-coordinates are distinct, and likewise all the $v$-coordinates are distinct. 
In this case we may, without loss, assume that the events are labelled so that the $u_{i}$ form a strictly increasing sequence. 
Then the causal relation of $S$ induces a total order on the $v$-coordinates, $v_i \leq v_j$. 
The product of the two total orders in $( u, v )$ is a 2D-order~\cite{2008BrightwellHensonSurya}. 
Non-degenerate sprinkles with equal cardinality that induce the same total order are necessarily isomorphic, but two non-degenerate sprinkles in the same isomorphism class can induce different orders. 
\begin{figure}
	\centering
	\includegraphics[scale=\graphicscaling]{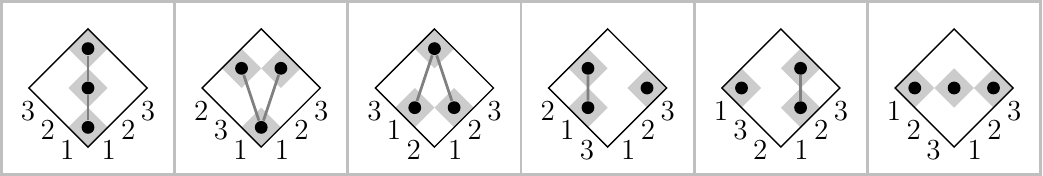}\\
	\includegraphics[scale=\graphicscaling]{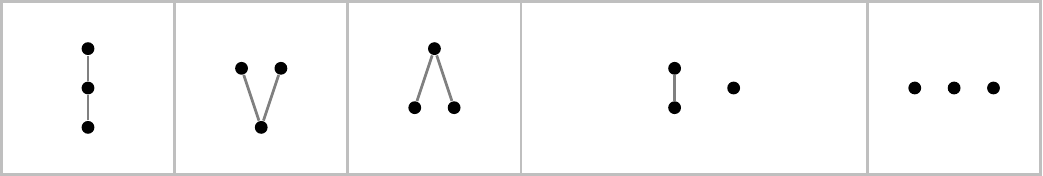}
	\caption{\label{fig:N3Posets} Choosing an event labelling by an increasing $u$-coordinate (upwards on the right of the diamonds), there exist 6 distinct total orders along the $v$-coordinate (upwards on the left of the diamonds). The corresponding 5 causets are shown at the bottom \cite{2020causets}.}
\end{figure}
This can be seen in \autoref{fig:N3Posets}, which displays an example of non-degenerate sprinkles inducing each of the 6 distinct total orders. 
Of these, there are two that are in the same isomorphism class, while the others correspond to distinct causets. 

For any finite causet $\mathscr{C}$, let $m( [ \mathscr{C} ] )$ be the number of total orders induced by non-degenerate sprinkles isomorphic to $\mathscr{C}$. 
In the case where $[ \mathscr{C} ]$ has no representative embedded in $1 + 1$ Minkowski, $m( [ \mathscr{C} ] ) = 0$; on the other hand, every causet that can be embedded in $1+1$ Minkowski can be embedded non-degenerately. 
Furthermore, a random sprinkle $\mathsf{S}$ in $1 + 1$ Minkowski spacetime is almost surely non-degenerate. 
While the combinatorics of random 2D-orders (including the number $m( [ \mathscr{C} ] )$) has been studied in the large cardinality limit~\cite{1990Winkler}, we want to use this idea to compute sprinkling probabilities for \emph{finite} causets in the following. 

\begin{proposition}
	\label{thm:PermutationProbability}
	Let $U$ be an Alexandrov subset of $1 + 1$ Minkowski spacetime. If $\mathscr{C}$ is a finite causet, the probability that a random sprinkle into $U$ with cardinality $|\mathscr{C}|$ has the same causal structure as $\mathscr{C}$ is
	\begin{align}
		\label{eq:PosetProbN3D2biconeNConditional}
				\Pr\bigl( 
					\mathsf{S} \sim \mathscr{C} 
				\bigm| 
					| \mathsf{S} | = | \mathscr{C} | 
				\bigr) 
		&= \frac{m\bigl( [ \mathscr{C} ] \bigr)}{| \mathscr{C} |!} 
		, 
	\end{align}
	Consequently,
	\begin{align}
		\nonumber
				\Pr\bigl( \mathsf{S} \sim \mathscr{C} \bigr) 
		&= \mu\bigl( [ \mathscr{C} ] \bigr) 
		\\\label{eq:PosetProbN3D2biconeN}
		&= \frac{m\bigl( [ \mathscr{C} ] \bigr)}{(| \mathscr{C} |!)^2} 
				\bigl( \rho \nu( U ) \bigr)^{| \mathscr{C} |} 
				e^{-\rho \nu( U )} 
		.
	\end{align} 
	\proof
	There are $n!$ distinct total orders of the $v$-coordinate for a given labelling of the $n$ event sprinkle. 
Each of these total orders has the same probability. 
Therefore, the probability for the random sprinkle $\mathsf{S}$ to be isomorphic to $\mathscr{C}$ given the cardinality $| \mathscr{C} |$ is the number $m( [ \mathscr{C} ] )$ of total orders along the $v$ coordinate that can be induced by any sprinkle $S \in [ \mathscr{C} ]$. 
	\qed
\end{proposition}

In the following, we use this combinatorial method to determine the expected size of the 1- and 2-layer past infinity of random sprinkles in Alexandrov subsets of $1 + 1$ dimensional Minkowski spacetime. 
We analytically approximate these probabilities, compare the analytic computations with results of simulations, and discuss the past infinity in the infinite causet limit. 

First, we make a general argument for random sprinkles $\mathsf{S}$ on any compact subset $U \in M$ of any given spacetime manifold $M$. 
Consider the canonical ensemble of sprinkles distributed according to the sprinkling measure but with fixed cardinality $n$. 
The expected cardinality of the 1-layer (index 1) or 2-layer (index 2) past infinity is given by a sum over the set $A( n )$ of all causets with cardinality $n$ (as shown in \autoref{tab:OEIS.A000112} for $n = 4$), 
\begin{align}
	\nonumber
			\mathbb{E}\bigl( 
				| C^{-}_{1, 2} | 
			\bigm| 
				| \mathsf{S} | = n 
			\bigr) 
	&= \sum_{\mathscr{C} \in A( n )} 
			\biggl( 
				\Pr\bigl( 
					\mathsf{S} \sim \mathscr{C} 
				\bigm| 
					| \mathsf{S} | = n 
				\bigr) 
	\\\label{eq:ExPastInfinityCanonical}
	&\quad\qquad\quad{}\times
				\bigl| C^{-}_{1, 2}( \mathscr{C} ) \bigr| 
			\biggr) 
	. 
\end{align}
Here $C^{-}_{1, 2}( \mathscr{C} )$ denotes the 1- or 2-layer past infinity of the causet $\mathscr{C}$. 
Because a general expression for the sets $A( n )$ is unknown, we use a different method to compute the expectation values. 

\begin{table*}
	\centering
	\begin{ruledtabular}
	\begin{tabular}{cl*{8}r}
		\multicolumn{2}{c}{} & 
		\multicolumn{8}{c}{$\mathbb{E}\bigr( | C^{-}_{1, 2} | \bigm| |\mathsf{S} | = n \bigl) / \%$} 
		\\
		\multicolumn{2}{r}{$n = $} 
		&       1 &       2 &       5 &      10 &      15 &      50 &     100 &     200 
		\\\hline
		\multirow{2}{*}{$C^{-}_{1}$}
		& analytic 
		&   100.0 &   75.00 &   45.67 &   29.29 &   22.12 &   8.998 &   5.187 &   2.939 
		\\
		& simulated
		&   100.0 &   74.96 &   45.61 &   29.28 &   22.09 &   8.986 &   5.191 &   2.941 
		\\\hline
		\multirow{2}{*}{$C^{-}_{2}$}
		& analytic 
		&   100.0 &   100.0 &   80.83 &   57.96 &   45.70 &   20.13 &   11.94 &   6.906 
		\\
		& simulated
		&   100.0 &   100.0 &   80.79 &   57.97 &   45.61 &   20.14 &   11.93 &   6.910 
		\\\hline\hline
		\multicolumn{2}{c}{} & 
		\multicolumn{8}{c}{$\mathbb{E}( | C^{-}_{1, 2} | ) / \%$} 
		\\
		\multicolumn{2}{r}{$\rho a^{2} = $} 
		&       1 &       2 &       5 &      10 &      15 &      50 &     100 &     200 
		\\\hline
		\multirow{2}{*}{$C^{-}_{1}$}
		& analytic 
		&   79.66 &   65.96 &   43.76 &   28.80 &   21.90 &   8.978 &   5.182 &   2.938 
		\\
		& simulated
		&   79.96 &   66.03 &   43.77 &   28.79 &   21.91 &   8.966 &   5.181 &   2.941 
		\\\hline
		\multirow{2}{*}{$C^{-}_{2}$}
		& analytic 
		&   97.82 &   93.08 &   76.43 &   56.63 &   45.10 &   20.08 &   11.92 &   6.902 
		\\
		& simulated
		&   97.96 &   93.27 &   76.49 &   56.59 &   45.10 &   20.05 &   11.92 &   6.910 
		\\
	\end{tabular}
	\end{ruledtabular}
	\caption{\label{tab:Ex1LayerPastInfinityD2bicone} Normalized expectation values for the 1-layer ($C^{-}_{1}$) or 2-layer ($C^{-}_{2}$) past infinity for ensembles with increasing causet cardinalities $n$ (top half), and increasing sprinkling density $\rho$ (bottom half) in units of the inverse volume $a^{-2}$, respectively. Simulated values are computed as averages over 100000 sprinkles.}
\end{table*}
An event sprinkled at position $x \in U$ is part of the 1-layer past infinity of a sprinkle with cardinality $n$ if all other $( n - 1 )$ events do not fall in its past region $U_{x} = J^{-}( x ) \cap U$ but appear in the remaining region $U \setminus J^{-}( x )$. 
For a fixed cardinality $n$ of a random sprinkle $\mathsf{S}$ on $U$, the expected size of the 1-layer past infinity (normalized by $n$) follows from the integral 
\begin{align}
	\nonumber
	&
			\frac{\mathbb{E}\bigl( 
					| C^{-}_{1} | 
				\bigm| 
					| \mathsf{S} | = n 
				\bigr) 
			}{n} 
	\\\label{eq:Ex1LayerPastInfinityCanonical}
	&= \frac{1}{\nu( U )^{n}} 
			\int_{U}
				\bigl( \nu( U ) - \nu( U_{x} ) \bigr)^{n - 1} 
			\id{\nu}( x ) 
	. 
\end{align}
Similarly, for the 2-layer past infinity, a sprinkled event at $x$ is in the 2-layer past infinity if it has any number $k \in \fieldN[0]$ of pairwise spacelike separated events to its past and the remaining $n - k - 1$ events are found again in the rest of the sprinkling region $U \setminus J^{-}( x )$. 
So the integral reads 
\begin{align}
	\nonumber
	&
			\frac{\mathbb{E}\bigl( 
					| C^{-}_{2} | 
				\bigm| 
					| \mathsf{S} | = n 
				\bigr) 
			}{n} 
	\\\nonumber
	&= \frac{1}{\nu( U )^{n}} 
			\sum_{k = 0}^{n - 1}
				\int_{U}
					\binom{n - 1}{k} 
					\bigl( \nu( U ) - \nu( U_{x} ) \bigr)^{n - 1 - k} 
	\\\label{eq:Ex2LayerPastInfinityCanonical}
	&\quad\qquad\qquad\qquad{}\times
					P_{k}( U_{x} ) 
					\bigl( \nu( U_{x} ) \bigr)^{k} 
				\id{\nu}( x ) 
	, 
\end{align}
where the weight $P_{k}( U_{x} )$ is the probability that the $k$ events form a subcauset $\mathscr{C}_{k, \dots, 1}$ of pairwise spacelike separated events within the region $U_{x}$. 
This probability is given by 
\begin{align}
	\nonumber
				P_{k}( U_{x} ) 
	&= \Pr\bigl( 
				\mathsf{S}_{x} \sim \mathscr{C}_{k, \dots, 1} 
			\bigm| 
				| \mathsf{S}_{x} | = k 
			\bigr) 
	\\\label{eq:ProbKEventsSpacelike}
	&= \frac{ 
				\mu_{U_{x}}\bigl( [ \mathscr{C}_{k, \dots, 1} ] \bigr) 
			}{ 
				\mu_{U_{x}}\bigl( Q_{U_{x}} \bigr) 
			}
\end{align}
where $\mathsf{S}_{x}$ is a random sprinkle on $U_{x}$. 

In the full grand canonical ensemble of all sprinkles, the cardinality $n$ is determined by the Poisson process with a fixed sprinkling density $\rho$, which leads to the normalized expectation values 
\begin{align}
	\nonumber
			\frac{ 
				\mathbb{E}\bigl( | C^{-}_{1, 2} | \bigr) 
			}{\rho \nu( U )} 
	&= \e^{- \rho \nu( U )} 
			\sum_{n = 1}^{\infty} 
			\Biggl( 
				\frac{\bigl( \rho \nu( U ) \bigr)^{n - 1}}{n!} 
	\\\label{eq:ExPastInfinityGrandCanonical}
	&\quad\qquad\qquad\quad{}\times
				\mathbb{E}\bigl( 
					| C^{-}_{1, 2} | 
				\bigm| 
					| \mathsf{S} | = n 
				\bigr) 
			\Biggr) 
	. 
\end{align}

The conditional expectation values (canonical ensemble) are easier to compute analytically and approximates the grand-canonical ensemble for larger sprinkling cardinalities, as we will see in the following. 

From these general results, we now return to the explicit computations in the special case of an Alexandrov subset $U$ in $1 + 1$ dimensional Minkowski spacetime. 
Here, we may express the position $x$ of an event in null coordinates $x = ( u', v' )$ ranging over $u', v' \in [ 0, a ]$ or rescaled over $u, v \in [ 0, 1 ]$ for a total volume of $\nu( U ) = a^2$, so that 
\begin{align}
	\label{eq:1LayerPastInfinityVolumeD2bicone}
			\nu( U_{x} )
	&= a^2 u v
	.
\end{align}
Evaluating \eqref{eq:Ex1LayerPastInfinityCanonical}, we find the normalized expected size of the 1-layer past infinity for cardinality $n$, 
\begin{align}
	\nonumber
	&
			\frac{\mathbb{E}\bigl( 
					| C^{-}_{1} | 
				\bigm| 
					| \mathsf{S} | = n 
				\bigr) 
			}{n} 
	= \frac{\mathrm{h}_{n}}{n} 
	\\\label{eq:Ex1LayerPastInfinityCanonicalFlat2D}
	&\asymp \frac{1}{n} 
		\left( 
			\ln( n ) 
		+ \upgamma 
		+ \frac{1}{2 n} 
		+ \mathcal{O}( n^{-2} ) 
		\right) 
	, 
\end{align}
where $\mathrm{h}_{n}$ is the $n$-th harmonic number, 
\begin{align}
	\label{eq:HarmonicNumber}
			\mathrm{h}_{n} 
	&:= \sum_{k = 1}^{n} \frac{1}{k} 
	. 
\end{align}
The asymptotic behavior for large $n$ is shown on the right hand side of \eqref{eq:Ex1LayerPastInfinityCanonicalFlat2D} where $\upgamma$ is the Eu\-ler-Ma\-sche\-ro\-ni constant. 
For large $n$, these asymptotics agree with the known results for random 2D-orders, for which the 1-layer past infinity is referred to as the minimal points of the partially ordered sets~\cite{1985Winkler,1990Winkler}. 
However, the following results are new. 

The normalized expectation value in the grand-canonical ensemble may be given in terms of the entire exponential integral $\Ein$, which is the generating function of harmonic numbers, 
\begin{align}
	\nonumber
			\frac{\mathbb{E}\bigl( | C^{-}_{1} | \bigr) 
			}{\rho a^2} 
	&= \frac{\Ein( \rho a^2 )}{\rho a^2} 
	\\\label{eq:Ex1LayerPastInfinityGrandCanonicalFlat2D}
	&= \frac{1}{\rho a^2} 
			\Bigl( 
				\ln( \rho a^2 ) 
			+ \upgamma 
			+ \Gamma( 0, \rho a^2 ) 
			\Bigr) 
	. 
\end{align}
The symbol $\Gamma( 0, z )$ is the incomplete Gamma function, which falls off rapidly in the limit $z \to \infty$, so that the asymptotic behavior (for $\rho \to \infty$) is the same as for the conditional expectation value \eqref{eq:Ex1LayerPastInfinityCanonicalFlat2D}. 

We complement these results by calculating the expected size of the 2-layer past infinity. 
Because the subset $U_{x} = J^{-}( x ) \cap U$ in \eqref{eq:ProbKEventsSpacelike} is an Alexandrov subset of $1 + 1$ dimensional Minkowski spacetime for all positions $x \in U$, the probability $P_{k}( U_{x} )$ is given by \autoref{thm:PermutationProbability} as the $x$-independent expression  
\begin{align}
	\label{eq:ProbKEventsSpacelikeFlat2D}
				P_{k}( U_{x} ) 
	&= \frac{1}{k!} 
	. 
\end{align}
There is only the total order $v_{k} < v_{k - 1} < \dotsb < v_{1}$ along the $v$-coordinate that corresponds to $k$ events being spacelike separated, assuming the $u$-coordinates are arranged in ascending order, $u_{1} < u_{2} < \dotsb < u_{k}$. 
Hence the integration in \eqref{eq:Ex2LayerPastInfinityCanonical} yields the expression 
\begin{align}
	\nonumber
	&
			\frac{\mathbb{E}\bigl( 
					| C^{-}_{2} | 
				\bigm| 
					| \mathsf{S} | = n 
				\bigr) 
			}{n} 
	= \frac{1}{n} 
			\sum_{k = 0}^{n - 1} \frac{\mathrm{h}_{n} - \mathrm{h}_{k}}{k!} 
	\\\label{eq:Prob2LayerPastInfinityCanonicalFlat2D}
	&\asymp \frac{\e}{n} 
		\left( 
			\ln( n ) 
		+ \tilde{\upgamma} 
		+ \frac{1}{2 n} 
		+ \mathcal{O}( n^{-2} ) 
		\right) 
\end{align}
with the constant 
\begin{align}
	\nonumber
			\tilde{\upgamma} 
	&= \upgamma - \frac{1}{\e} \sum_{k = 0}^{\infty} \frac{\mathrm{h}_{k}}{k!} 
	\\\nonumber
	&= \upgamma - \Ein( 1 ) 
	\\\label{eq:Prob2LayerPastInfinityAsymptoticConstant}
	&\approx -0.21938 
	. 
\end{align}
We do not have an expression for the expectation value in the grand-canonical case, however, the summation in \eqref{eq:ExPastInfinityGrandCanonical} is quickly converging so that it can be computed numerically with sufficient accuracy. 

\autoref{tab:Ex1LayerPastInfinityD2bicone} shows some examples for the normalized expectation values at fixed cardinalities $n$ (canonical) and fixed sprinkling densities $\rho$ (grand-canonical). 
The numbers are presented as a percentage, since they may also be interpreted as the probabilities that an event randomly chosen from a sprinkle is in the 1- or 2-layer past infinity, respectively. 
The simulations results below the analytic results are computed from the cardinalities of the past infinities averaged over 100000 sprinkles. 
Note that the values for the two ensembles become asymptotically equal as the cardinality increases. 
Furthermore, the values decrease with increasing sprinkle cardinality so that sufficiently many events in sprinkles of more than 200 events lie outside the 2-layer past infinity and thus have non-empty rank 2 pasts. 
As the sprinkle size increases, the proportion of the sprinkle lying in past infinity tends to zero and the influence of the past infinity on the preferred pasts becomes negligible. 

\begin{remark}
	Normalized expected sizes of higher $j$-layer past infinities (with $j \in \fieldN$) are computed with the same integral \eqref{eq:Ex2LayerPastInfinityCanonical} as for the 2-layer past infinity. 
	However, in general we have to account for all possible arrangements of the $k$ events to the past of position $x \in U$ such that an event at $x$ is part of the $j$-layer past infinity. 
	Thus the probability weight \eqref{eq:ProbKEventsSpacelike} has to be replaced by the sum  
	\begin{align}
		\label{eq:ProbKEventsHigherLayerPastInfinity}
				P^{(j)}_{k}( U_{x} ) 
		&= \sum_{\mathscr{C}_{j}} 
				\Pr\bigl( 
					\mathsf{S}_{x} \sim \mathscr{C}_{j} 
				\bigm| 
					| \mathsf{S}_{x} | = k 
				\bigr) 
	\end{align}
	running over all causets $\mathscr{C}_{j}$ with cardinality $k$ such that an event that has a sprinkle isomorphic to $\mathscr{C}_{j}$ as its past is in the $j$-layer past infinity. 
	All subcausets $\mathscr{C}_{j}$ can have at most $j - 1$ layers.
	For $j = 1$, the sum is trivially 1 since $k = 0$, and for $j = 2$, there is only one term, the $k$-event \emph{antichain}.
\end{remark}

\section{Conclusion}
\label{sec:Conclusion}
In this work, we used ensembles of 10000 sprinkles in Alexandrov subsets of $1 + 1$ to $1 + 3$ dimensional Minkowski spacetime to study the preferred past structure for causal sets, which was recently proposed as a supplement to causets in order to discretize the Klein-Gordon field equation~\cite{2020DableheathEtAl}. 
We compared 6 criteria to find subsets of the rank 2 past that can be used to determine a preferred past by investigating the corresponding past diamonds. 
As criterion 1, we considered the largest diamond criterion that was suggested in~\cite{2020DableheathEtAl}, which performs well in selecting a unique event of the rank 2 past for almost every event in a causet. 
However, since the diamonds according to this criterion can be arbitrarily large, the proper time distribution of the diamonds has a large expectation value. 
It transpired that criterion 6 performs best in selecting a unique diamond with the highest probability among the investigated criteria. 
The distribution of the proper time separation for the diamonds selected by criterion 6 has a relatively small expectation value and small variance. 
The selected rank 2 past events are approximately uniformly distributed on the unit past hyperboloid, which indicates that criterion 6 tends to be Lorentz invariant in the large limit. 
For criterion 6, we first minimize the number of internal events and then maximize the number of perimetral events among those past diamonds that are unique (have no duplicate among the diamonds in the rank 2 past). 
If there is no singleton rank 2 past subset selected by this rule, then criterion 6 takes the subset of rank 2 past events spanning diamonds that minimize the number of internal events among those diamonds with a maximal number of perimetral events. 

We also analysed the diamonds that are spanned by next-to-nearest neighbours along geodesic paths through the sprinkled causets. 
These diamonds are always pure, mostly small and their distribution is similar across the three considered flat spacetime dimensions. 
This observation supports the argument of a dimensional reduction for small causal intervals in causets~\cite{2015Carlip}. 
One might hope that a discretization method for field equations on causal sets should be independent of the spacetime dimension, which is an emergent property rather than built in as a fundamental parameter. 
Therefore, the indication of dimensional independence of small diamonds tends to support the use of criteria that select such diamonds, like criterion 6.
Within the limits of our numerical investigations, we conclude that a preferred past structure determined by criterion 6 may give rise to a dimension independent discretization method. 
Further studies with an explicit comparison to the spacetime continuum and investigations of sprinkles on curved spacetimes are open tasks. 

In the second part of the work, we reviewed the construction of the Poisson probability measure~\cite{1998AlbeverioKondratievRoeckner} and applied the results in the context of sprinklings in causal set theory for a given spacetime manifold. 
The sprinkling probability space can be used to discuss the limit to infinite causal sets. 
For now, we used this method to determine the expected size of the 1- and 2-layer past infinities for Alexandrov subsets of $1 + 1$ dimensional Minkowski spacetime normalized by the causet cardinality. 
This served as a consistency check for the numerical analysis. 
We showed that in the limit of arbitrarily large causets the proportionate size of the past infinity is negligible. 

In general, the rank 2 past of an event in an infinite causet is infinite and we do not expect that any of the criteria presented above would still select a singleton subset with high probability.
However, this problem could be avoided by working on a past-finite subset, which is the analogue of a past-compact subset in the continuum.
In globally hyperbolic spacetimes, the future of a Cauchy surface is a past-compact subset. 
The definitions of the causet analogues of a globally hyperbolic spacetime and Cauchy surface may need to be refined so that a similar statement is true. 

Potential applications of our results include the algebraic formulation of (quantum) fields on causal sets and the study of the causet analogues of Cauchy surfaces. 
They could also be adapted to study preferred past structures for sprinkles on curved spacetimes. 
The probability space for sprinkling into arbitrary globally hyperbolic spacetimes given in the second part of this paper may facilitate a more general discussion of the continuum limit of causal sets.
We leave these ideas for future projects. 

\begin{acknowledgments}
CM would like to thank Stav Zalel and Ian Jubb for helpful remarks and literature suggestions, as well as everyone who participated in enriching discussions at the virtual conferences ``Quantum Gravity 2020'' and ``The Virtual Causet''. 
Alexei Daletskii suggested to review the publication~\cite{1998AlbeverioKondratievRoeckner} for the construction of the Poisson probability measure. 
Fleur Versteegen and Nomaan X have shared their experiences when setting up the numerical simulations. 

CM received an EPSRC funding (grant number EP/N509802/1) for his PhD fellowship that made this research possible. 

The simulations in this project were undertaken on the Viking Cluster, which is a high performance compute facility provided by the University of York. 
We are grateful for computational support from the University of York High Performance Computing service, Viking and the Research Computing team.
\end{acknowledgments}

\appendix
\section{Implementation of the sprinkling process}
\label{sec:AppendixSprinklingImplementation}
The sprinkling process is implemented as follows. 
The sprinkling region is an Alexandrov interval $U$ in $d$-dimensional Minkowski spacetime. 
Using Lorentz invariance of the sprinkling measure, we may, without loss, take $U$ to be the Cauchy development of a ball of radius $R$ centred a the origin of the $t = 0$ hyper-surface in standard inertial coordinates. 

A sprinkle is obtained as follows. 
\begin{enumerate}
  \item Randomly choose the sprinkle cardinality $n \in \fieldN[0]$ according to the Poisson distribution with mean 6000. 
  \item The sprinkle comprises $n$ events, each of which has a spacetime position $x$ chosen independently from a uniform distribution on $U$ w.r.t.\ the volume measure. 
  	This is achieved by setting 
  \begin{align*}
				x 
		&= \left( t_{\mathrm{sign}} ( 1 - h ) R, r R \frac{\vec{v}}{| \vec{v} |} \right) 
		,
	\end{align*}
  	where:
		\begin{enumerate}
		  \item $t_{\mathrm{sign}} \in \{ -1, 1 \}$ is the uniformly chosen sign of the time coordinate.
		  \item $h = u_{h}^{1 / d}$ is determined by a uniformly distributed random number $u_{h} \in [0, 1]$. 
		  \item $r = h u_{r}^{1 / (d - 1)}$ is the radial scaling determined by a uniformly distributed random value $u_{r} \in [0, 1]$. 
		  \item $\vec{v} \in \fieldR^{d - 1}$ is a vector with components that are independently chosen from a normal distribution with zero mean and unit variance, such that the resulting normalised vectors $\vec{v} / | \vec{v} |$ are uniformly distributed on the unit $( d - 2 )$-sphere. 
		\end{enumerate}
\end{enumerate}

\section{Cardinality of the rank 2 past}
\label{sec:AppendixRank2Past}
\begin{figure}
	\centering
	\includegraphics[scale=\graphicscaling]{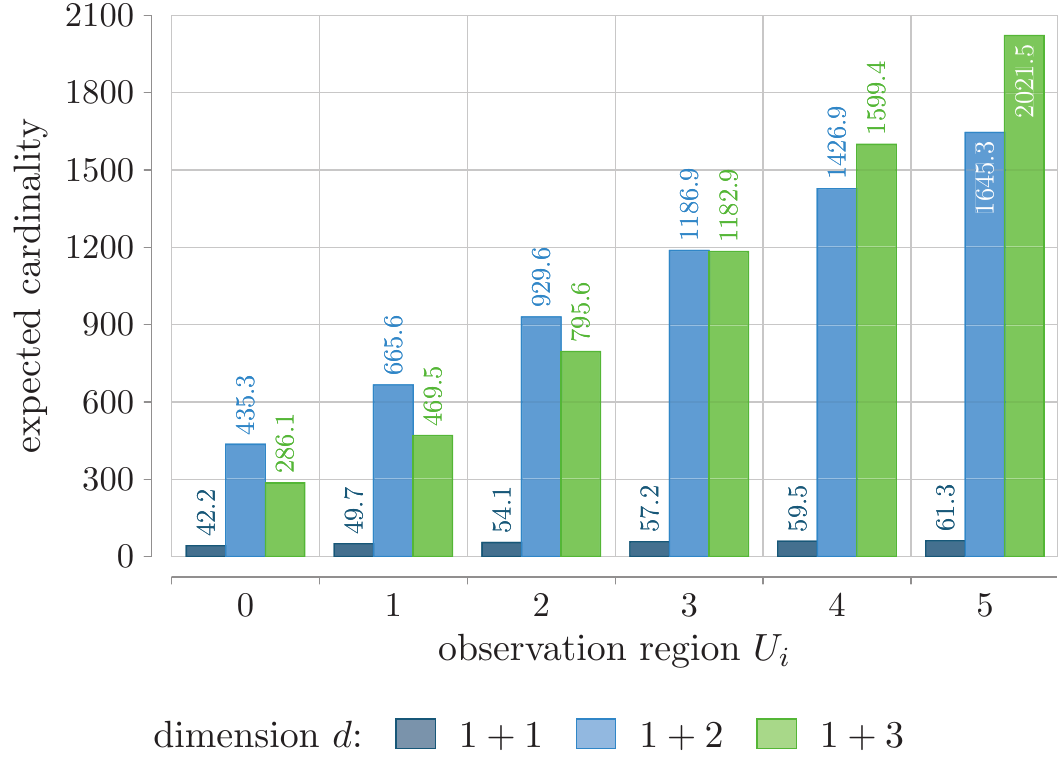}
	\caption{\label{fig:Rank2Past} Discrete distributions of the expected numbers of rank 2 past events for a random event $\mathsf{x}$ that is not in the 2-layer past infinity of a sprinkle on the Alexandrov subset $U$ of $1 + 1$ (darkest blue shade), $1 + 2$ (lighter blue shade), and $1 + 3$ (lightest/green shade) dimensional Minkowski spacetime. For all dimensions, the expected cardinality of $R^{-}_{2}( \mathsf{x} )$ increases with the observation region $U_{i}$ from $i = 0$ to $i = 5$.}
\end{figure}
In this appendix, we show the cardinality of the rank 2 past for a typical event in our sprinkles on an Alexandrov subset $U$ of $1 + 1$, $1 + 2$, or $1 + 3$ dimensional Minkowski spacetimes. 
The sprinkles are generated by a Poisson process with an expected total cardinality of 6000 events. 

For a given causet $\mathscr{C}$ in the ensemble of sprinkles, a random event $\mathsf{x} \in \mathscr{C} \setminus C^{-}_{2}$ (not in the 2-layer past infinity) has a rank 2 past with an expected cardinality as displayed in \autoref{fig:Rank2Past}. 
We can see that the expected cardinality of the rank 2 past $R^{-}_{2}( \mathsf{x} )$ grows with the cardinality of the past $J^{-}( \mathsf{x} )$, since the past of $\mathsf{x}$ becomes larger with decreasing volume (increasing index $i$) of the observation region $U_{i}$. 

In arbitrary large sprinkles, this growth is unbounded and events in infinite causal sets typically have infinitely many links to their past, thus also infinitely many elements in the rank 2 past. 

\section{Non-empty subsets selected by the 6 criteria}
\label{sec:AppendixRank2PastCriteria}
It remains to show that the rank 2 past subsets selected by our 6 criteria are non-empty. 
For the proofs that any of our criteria yields a non-empty subset of the rank 2 past, note that a causet event that is not part of the 2-layer past infinity has a non-empty rank 2 past, so we can make the following arguments. 

\begin{lemma}
	\label{lma:NonEmptyPureDiamonds}
	If $x \in \mathscr{C} \setminus C^{-}_{2}$ for some causet $\mathscr{C}$, then $R^{-}_{2}( x )$ contains at least one event that spans a pure diamond with $x$. 
	
	\proof
	Take any event $y_{0} \in R^{-}_{2}( x )$. 
	Either $[ y_0, x ]$ is pure or it contains internal events including at least one event $y_{1}$ that is also two links in the past of $x$, $y_{1} \in R^{-}_{2}( x )$. 
	The diamond $[ y_{1}, x ] \subset [ y_{0}, x ]$ is either pure or contains yet another internal event that is also in the rank 2 past, $y_{2} \in R^{-}_{2}( x )$. 
	This process may be repeated until it terminates (recall that $[ y, x ]$ is finite) with a diamond spanned by $x$ and an event $y_{i} \in R^{-}_{2}( x )$ for some $i \in \fieldN[0]$ such that $[ y_{i}, x ]$ has no internal events (it is pure). 
	\qed
\end{lemma}
For example, the two smallest diamonds (the 1- and the 2-diamond) are pure. 
Out of the two possible 3-diamonds, one is pure and the other contains a 1-diamond, and so on. 
\begin{proposition}
	Let $x \in \mathscr{C} \setminus C^{-}_{2}$ for some causet $\mathscr{C}$.  
	All subsets of its rank 2 past that are determined by the six criteria (defined in Sec.~\ref{sec:Rank2PastCriteria}) are non-empty. 
	
	\proof
	Non-emptiness of the subsets for criteria 1 (largest diamonds) and 2 (smallest diamonds) is a direct consequence of the fact that the functions 
	\begin{align*}
				\argmax_{y \in R^{-}_{2}( x )} \bigl| [ y, x ] \bigr| 
		&\qquad\text{and}\qquad 
				\argmin_{y \in R^{-}_{2}( x )} \bigl| [ y, x ] \bigr| 
	\end{align*}
	are taken over the non-empty set $R^{-}_{2}( x )$. 
	For criterion 3 (largest pure diamonds), we consider the subset of pure diamonds only, which is non-empty as shown in \autoref{lma:NonEmptyPureDiamonds}, so that the $\argmax$ function yields again a non-empty subset. 
	Criteria 4 and criteria 5 take the extrema of two properties in succession, so that their selections are non-empty. 
	Finally, criterion 6 yields either a singleton or the same result as criterion 5 if there are no singletons among all subsets $D^{-}_{i, p}( x )$ as defined in \eqref{eq:DiamondsSetMatrix}. 
	Any singleton is by definition non-empty and we have just shown that the rank 2 past subset given by criterion 5 is non-empty as well. 
	So in summary, all criteria yield a non-empty subset of rank 2 events for any causet event that has a non-empty rank 2 past. 
	\qed
\end{proposition}

\bibliographystyle{apsrev4-2}

\begin{thebibliography}{99}

\bibitem{1987BombelliEtAl}
	L.~Bombelli, J.~Lee, D.~Meyer, and R.~D.~Sorkin,
	Space-time as a causal set,
	\href{https://doi.org/10.1103/PhysRevLett.59.521}{Phys.\ Rev.\ Lett., \textbf{59}, 521 (1987)}.

\bibitem{2009Henson}
	J.~Henson,
	The causal set approach to quantum gravity,
	in \textit{Approaches to Quantum Gravity}, edited by D.~Oriti (Cambridge University Press, Cambridge, England, 2009) pp.~393ff.

\bibitem{2011Sorkin}
	R.~D.~Sorkin,
	Scalar field theory on a causal set in histories form,
	\href{https://doi.org/10.1088/1742-6596/306/1/012017}{J.\ Phys.\ Conf.\ Ser.\ \textbf{306}, 012017 (2011)}.

\bibitem{2014AslanbeigiSaravaniSorkin}
	S.~Aslanbeigi, M.~Saravani, and R.~D.~Sorkin,
	Generalized causal set d'Alembertians,
	\href{https://doi.org/10.1007/JHEP06(2014)024}{J.\ High Energy Phys. 06 (2014), 024}.

\bibitem{2013DowkerGlaser}
	F. Dowker and L. Glaser,
	Causal set d'Alembertians for various dimensions,
	\href{https://doi.org/10.1088/0264-9381/30/19/195016}{Classical Quantum Gravity, \textbf{30}, 195016 (2013)}.

\bibitem{2014Glaser}
	L.~Glaser,
	A closed form expression for the causal set d'Alembertian,
	\href{https://doi.org/10.1088/0264-9381/31/9/095007}{Classical Quantum Gravity \textbf{31}, 095007 (2014)}.

\bibitem{2020DableheathEtAl}
	E.~Dable-Heath, C.~J.~Fewster, K.~Rejzner, and N.~Woods,
	Algebraic classical and quantum field theory on causal sets,
	\href{https://doi.org/10.1103/PhysRevD.101.065013}{Phys.\ Rev.\ D \textbf{101}, 065013 (2020)}.

\bibitem{2009Sorkin}
	R.~D.~Sorkin,
	Does locality fail at intermediate length-scales,
	in \textit{Approaches to Quantum Gravity}, edited by D.~Oriti (Cambridge University Press, Cambridge, England, 2009), pp.~26--43.

\bibitem{1998AlbeverioKondratievRoeckner}
	S.~Albeverio, Y.~G.~Kondratiev, and M.~R\"ockner,
	Analysis and geometry on configuration spaces,
	\href{https://doi.org/10.1006/jfan.1997.3183}{J.\ Funct.\ Anal.\ \textbf{154}, 444 (1998)}.

\bibitem{2008BrightwellHensonSurya}
	G.~Brightwell, J.~Henson, and S.~Surya,
	A 2D model of causal set quantum gravity: The emergence of the continuum,
	\href{https://doi.org/10.1088/0264-9381/25/10/105025}{Classical Quantum Gravity \textbf{25}, 105025 (2008)}.

\bibitem{1990Winkler}
	P.~Winkler,
	Random orders of dimension 2,
	\href{https://doi.org/10.1007/BF00383197}{Order \textbf{7}, 329 (1990)}.

\bibitem{1988Meyer}
	D.~A.~Meyer,
	The dimension of causal sets, Ph.~D.\ thesis, 
	Massachusetts Institute of Technology, 1988.

\bibitem{1978Myrheim}
	J.~Myrheim,
	Statistical geometry, European Organization for Nuclear Research (CERN) 
	Technical Report No.~2538, 1978.

\bibitem{2003Reid}
	D.~D.~Reid,
	Manifold dimension of a causal set: Tests in conformally flat spacetimes,
	\href{https://doi.org/10.1103/PhysRevD.67.024034}{Phys.\ Rev.\ D \textbf{67}, 024034 (2003)}.

\bibitem{2013RoySinhaSurya}
	M.~Roy, D.~Sinha, and S.~Surya,
	Discrete geometry of a small causal diamond,
	\href{https://doi.org/10.1103/PhysRevD.87.044046}{Phys.\ Rev.\ D \textbf{87}, 044046 (2013)}.

\bibitem{2015Carlip}
	S.~Carlip,
	Dimensional reduction in causal set gravity,
	\href{https://doi.org/10.1088/0264-9381/32/23/232001}{Classical Quantum Gravity \textbf{32}, 232001 (2015)}.

\bibitem{2000Bombelli}
	L.~Bombelli,
	Statistical Lorentzian geometry and the closeness of Lorentzian manifolds,
	\href{https://doi.org/10.1063/1.1288494}{J.\ of Math.\ Phys.\ (N.Y.) \textbf{41}, 6944 (2000)}.

\bibitem{1999AlbeverioKondratievRoeckner}
	S.~Albeverio, Y.~G.~Kondratiev, and M.~R{\"o}ckner,
	Diffeomorphism groups and current algebras: configuration space analysis in quantum theory,
	\href{https://doi.org/10.1142/S0129055X99000027}{Rev.\ Math.\ Phys.\ \textbf{11}, 1 (1999)}.

\bibitem{2005Parthasarathy}
	K.~R.~Parthasarathy,
	\textit{Probability Measures on Metric Spaces}, 
	\href{https://doi.org/10.1090/chel/352}{(American Mathematical Society, Providence, 1967), Vol.~352}.

\bibitem{website:oeisA000112}
	OEIS Foundation Inc.,
	The on-line encyclopedia of integer sequences,
	\href{http://oeis.org/A000112}{http://oeis.org/A000112} (2020).

\bibitem{2002BrinkmannMcKay}
	G.~Brinkmann and B.~D.~McKay,
	Posets on up to 16 Points,
	\href{https://doi.org/10.1023/A:1016543307592}{Order \textbf{19}, 147 (2002)}.

\bibitem{1985Winkler}
	P.~Winkler,
	Random orders,
	\href{https://doi.org/10.1007/BF00582738}{Order \textbf{1}, 317 (1985)}.

\bibitem{2020causets}
	C.~Minz, Causets--Draw causal set (Hasse) diagrams, 
	\href{https://ctan.org/pkg/causets}{https://ctan.org/pkg/causets} (2020). 

\end{thebibliography}

\end{document}